\documentclass[12pt]{article}
\usepackage{epsf}
\usepackage{amsmath,amssymb}
\usepackage[dvipdfmx]{graphicx}
\usepackage{color}
\usepackage{cite}
\usepackage[colorlinks,citecolor=blue]{hyperref}
\renewcommand{\thefootnote}{\#\arabic{footnote}}

\renewcommand{\thefootnote}{\fnsymbol{footnote}}
\def\thefootnote{\fnsymbol{footnote}}

\newcommand{\Mpl}{M_{\rm pl}}
\newcommand{\al}{\alpha}
\newcommand{\ep}{\epsilon}
\newcommand{\pr}{\prime}

\makeatletter

\@addtoreset{equation}{section}
\makeatother

\begin{document}

\begin{titlepage}

\begin{center}

\vskip .75in

{\Large \bf Relaxing inflation models with non-minimal coupling: \\ A general study}

\vskip .75in

{\large
Tatsuki~Kodama$\,^1$ and  Tomo~Takahashi$\,^2$ 
}

\vskip 0.25in

{\em
$^{1}$Graduate School of Science and Engineering, Saga University, Saga 840-8502, Japan
\vspace{2mm} \\
$^{2}$Department of Physics, Saga University, Saga 840-8502, Japan
}

\end{center}
\vskip .5in

\begin{abstract}

We investigate the predictions of inflation models with a non-minimal coupling to gravity for  inflationary observables such as the spectral index and tensor-to-scalar ratio  in a general setting. We argue that, depending on the relation between the Jordan frame potential  and a function characterizing the non-minimal coupling, one can classify the model into three categories each of which gives distinctive predictions for the inflationary observables.  We derive general predictions for each class and also investigate some explicit models to discuss how the general features arise.  Our results would be useful to design an inflation model consistent with observational constraints with a non-minimal coupling to gravity.

\end{abstract}

\end{titlepage}

\renewcommand{\thepage}{\arabic{page}}
\setcounter{page}{1}
\renewcommand{\thefootnote}{\#\arabic{footnote}}
\setcounter{footnote}{0}

\section{Introduction \label{sec:intro}}

Inflation has been a successful paradigm to describe the very early stage of the  Universe. The inflationary expansion is considered to be driven by a scalar field, called the inflaton, and its perturbations provide primordial density fluctuations through which we can probe the inflationary epoch by observing anisotropies of cosmic microwave background (CMB), large scale structure, and so on. Indeed, current cosmological observations of CMB such as Planck satellite \cite{Planck:2018jri}  have severely constrained the so-called spectral index $n_s$ and the tensor-to-scalar ratio $r$,  now critically test inflation models. In particular, recent results from CMB B-mode polarization experiment of BIECP/Keck 2018 \cite{BICEP:2021xfz}, in combination with Planck data,  have put a more  stringent constraint on $n_s$ and $r$, which now excluded chaotic inflation with any power-law potential and natural inflation models  \cite{BICEP:2021xfz}.

Although these inflation models  are  disfavored by current observational constraints as a model minimally coupled to gravity, if one introduces a non-minimal coupling, the inflationary prediction can be modified and may become consistent with observations. One of such a famous example is the Higgs inflation \cite{Bezrukov:2007ep} (see also \cite{Spokoiny:1984bd,Futamase:1987ua,Salopek:1988qh,Fakir:1990eg,Komatsu:1999mt} for earlier works of this type), in which quartic chaotic inflation  has a non-minimal coupling to gravity. Although its minimally coupled version has already been excluded due to too large tensor-to-scalar ratio, it becomes a viable model when one includes a non-minimal coupling mainly because the tensor-to-scalar ratio is suppressed in such models \cite{Komatsu:1999mt}.  By introducing a non-minimal coupling, the predictions for $n_s$ and $r$ are in general modified and one may be able to relax inflation models like in the case of the Higgs inflation. Such examples include quadratic chaotic inflation \cite{Linde:2011nh,Boubekeur:2015xza,Tenkanen:2017jih}, power-law inflation and inverse monomial inflation \cite{Takahashi:2020car},  natural inflation \cite{Ferreira:2018nav},  and so on\footnote{
One can also consider a multi-field extension to alleviate models which are excluded by observational constraints. For example, spectator field models such as the curvaton \cite{Enqvist:2001zp,Lyth:2001nq,Moroi:2001ct}, modulated reheating  scenario \cite{Dvali:2003em,Kofman:2003nx} and so on in which another scalar field other than the inflaton, whose energy density is negligible during inflation,  generates primordial fluctuations, whereas the inflationary expansion is driven by the inflaton. In such a model, the predictions for the inflationary parameters are also modified, and interestingly, some inflation models can becomes viable even though it is excluded as a single-field inflation model \cite{Langlois:2004nn,Moroi:2005kz,Moroi:2005np,Ichikawa:2008iq,Ichikawa:2008ne,Enqvist:2013paa,Vennin:2015vfa,Haba:2017fbi}.
}.

However, one should be cautioned that a non-minimal coupling does not necessarily modify  inflation models in such a way that the model becomes viable against cosmological data. For instance, when a negative non-minimal coupling is assumed for a quadratic chaotic inflation model, the tensor-to-scalar ratio gets enhanced and it is inconsistent with the data \cite{Boubekeur:2015xza}.  It also happens that if one assumes too large non-minimal coupling, the spectral index and/or the tensor-to-scalar ratio are modified  much and the model predictions move away from the observationally allowed region.  Therefore it is not trivial whether a model gets resurrected or not by assuming a non-minimal coupling. Furthermore, even though many inflation models (inflaton potential) have been proposed and studied in the context of the single-field inflation framework,  only a handful of models have been investigated in the  non-minimally coupled version. Besides, the analysis for non-minimal inflation so far is mostly model-dependent and  a systematic study has been scarcely done regarding what inflation model (inflaton potential) and what kind of non-minimal coupling give successful predictions for the inflationary observables such as $n_s$ and $r$.  

In light of these considerations, it would be worthwhile to investigate general predictions for inflationary observables in non-minimal inflation models systematically and as model-independently as possible, which is the main purpose of this paper. Based on a general setting, we study the inflationary predictions  to see what kind of non-minimal coupling can relax inflation models even if they are excluded as a minimally coupled model.

The structure of this paper is as follows. In the next section, we describe the framework we consider in this paper and give expressions for the inflationary observables in a model-independent manner. To discuss how the predictions of inflation models are modified due to the existence of a non-minimal coupling, we classify models into three categories by using the relation between the Jordan frame inflaton potential  and a function specifying how the inflaton is non-minimally coupled to gravity.   Then, in Section~\ref{sec:predictions}, we argue the general predictions of non-minimal inflation models and how one can relax the minimally coupled counterpart.  The final section is devoted to discussion and conclusion of this paper.

\section{Framework \label{sec:framework}}

\subsection{General non-minimal inflation}

Here we describe the framework to study the predictions for inflationary observables in a general setting. We consider a single-field inflation model non-minimally coupled to gravity whose  Jordan frame action is given as\footnote{
One can also include a functional uncertainty  for the kinetic term of $\phi$ as 
$ \frac{1}{2}  g^{\mu\nu} \left( \partial_{\mu}\phi  \right) \left( \partial_{\nu}\phi \right)
\to 
\frac{1}{2} B(\phi)  g^{\mu\nu} \left( \partial_{\mu}\phi  \right) \left( \partial_{\nu}\phi \right)
$, which has been discussed in \cite{Jarv:2016sow,Jarv:2014hma}. However, the effects of $B(\phi)$ is somewhat degenerate with those of $A(\phi)$ after the Weyl transformation,  and hence we do not consider such additional freedom for the kinetic term in this paper.
} 
\begin{equation}
 \label{nonminimal_action_J}
S_{\rm J} = 
\int d^4x \sqrt{-g}\left(-\frac{1}{2} A(\phi) \Mpl^2 \, g^{\mu\nu}R_{\mu\nu}(\Gamma) 
+ \frac{1}{2}  g^{\mu\nu} \left( \partial_{\mu}\phi  \right) \left( \partial_{\nu}\phi \right) - V_J(\phi) \right)  \,,
\end{equation}
where $\phi$ is an inflaton, $A(\phi)$ is a function representing a non-minimal coupling to gravity and 
$V_J(\phi)$ is the Jordan frame potential for $\phi$.   

By making a Weyl transformation
\begin{equation}
\label{ }
\hat{g}_{\mu \nu} =  \Omega^2 (\phi) g_{\mu\nu} \,, 
\qquad
\Omega^2 (\phi) = A(\phi)  \,, 
\end{equation}
the action can be brought into the Einstein frame one: 
\begin{equation}
 \label{nonminimal_action_E}
S_{\rm E} = 
\int d^4x \sqrt{-\hat{g}}\left(-\frac{1}{2}  \Mpl^2 \, \hat{g}^{\mu\nu} \, \hat{R}_{\mu\nu}(\hat{\Gamma}) 
+ \frac{1}{2}  \hat{g}^{\mu\nu} \left( \partial_{\mu} \chi  \right) \left( \partial_{\nu}\chi \right) - V_E(\chi (\phi) ) \right)  \,,
\end{equation}
in which the quantities with a hat represent the one in the Einstein frame and $\chi$ is the Einstein frame field related to $\phi$ as 
\begin{equation}
\frac{d\phi}{d \chi} = \sqrt{\frac{2A(\phi)^2}{2 A (\phi) + 3 \kappa \Mpl^2 \left(A'(\phi) \right)^2}} \,.
\end{equation}
Here a prime represents the derivative with respect to $\phi$, i.e., $A'(\phi) = dA /d\phi$. Here $\kappa =1$ corresponds to the case of the metric formulation, while $\kappa=0$ is for the Palatini case\footnote{
See, e.g., \cite{Tenkanen:2020dge} for an introductory review of the Palatini formulation.
}.
$V_E (\chi(\phi))$ is the potential in the Einstein frame which can be written as 
\begin{equation}
\label{eq:V_E}
V_E (\chi(\phi)) = \frac{V_J(\phi)}{A^2 (\phi)} = \frac{V_J(\phi)}{\Omega^4 (\phi)}  \,.
\end{equation}

To predict the inflationary observables such as the spectral index $n_s$ and the tensor-to-scalar ratio $r$, we need to compute the slow-roll parameters, which are defined in the Einstein frame as 
\begin{equation}
\epsilon = \frac12 \Mpl^2 \left( \frac{1}{V_E} \frac{dV_E}{d\chi} \right)^2, 
\qquad
\eta = \Mpl^2 \frac{1}{V_E} \frac{d^2V_E}{d\chi^2} \,,
\end{equation}
with which the spectral index $n_s$ and the tensor-to-scalar ratio are given by 
\begin{equation}
\label{ }
n_s = 1 - 6 \epsilon +2 \eta \,,
\qquad
r = 16 \epsilon \,.
\end{equation}
We also sometimes use the slow-roll parameters defined for the Jordan frame potential, denoted as $\epsilon_J$ and $\eta_J$, which are given by 
\begin{equation}
\epsilon_J  = \frac12 M_{\rm pl}^2 \left( \frac{1}{V_J} \frac{d V_J}{d \phi} \right)^2,
\qquad\qquad
\eta_J =  M_{\rm pl}^2  \frac{1}{V_J} \frac{d^2 V_J}{d \phi^2} \,.
\end{equation}
By using $\epsilon_J$ and $\eta_J$,  we can express the Einstein frame slow-roll parameters $\epsilon$ and $\eta$ as 
\begin{eqnarray}
\label{eq:eps_general}
&& \epsilon = 
P^2 \left[ 
\epsilon_J  - 2 M_{\rm pl}^2 \frac{A'}{A} \frac{V_J'}{V_J} + 2 M_{\rm pl}^2  \left( \frac{A'}{A} \right)^2 
\right], \\ 
\label{eq:eta_general}
&& \eta = 
P^2 
\left[ 
\eta_J 
- 4 M_{\rm pl}^2 \frac{A'}{A} \frac{V_J'}{V_J}
+ 6 M_{\rm pl}^2  \left( \frac{A'}{A} \right)^2 
- 2  M_{\rm pl}^2   \frac{A^{''}}{A} 
+   M_{\rm pl}^2 \frac{P'}{P}\left( - 2 \frac{A'}{A}  + \frac{V_J'}{V_J} \right)
\right] \,,  \notag \\
\end{eqnarray}
where we have defined $P= P(\phi) = d\phi / d\chi$. The difference between the metric and Palatini formulation only appears in the functional expression of $P(\phi)$, and hence Eqs.~\eqref{eq:eps_general} and \eqref{eq:eta_general}  hold  for both formulations. 

Furthermore, one often provides the analytic formulas for $n_s$ and $r$ in terms of the number of $e$-folds $N$, which can be calculated as 
\begin{equation}
\label{eq:N_general}
N = \frac{1}{\Mpl^2} \int_{\phi_{\rm end}}^{\phi_\ast}  \frac{1}{P^2} \left( \frac{V_J'}{V_J} - 2  \frac{A'}{A}  \right)^{-1} d\phi \,,
\end{equation}
where $\phi_\ast$ and $\phi_{\rm end}$ are the values of $\phi$ at which the reference scale exited the horizon during inflation and the end of inflation, respectively. 

Up to here, the expressions are quite general.  Since the functional form of $A(\phi)$ assumed in most works of non-minimal inflation is given as  
\begin{equation}
\label{eq:A_phi}
 A(\phi)  = 1 + \xi  {\cal F}(\phi) \,, 
\end{equation}
with $\xi$ being the dimensionless coupling parameter, we take the above form for $A(\phi)$ in the rest of this paper. Indeed in many works,   $A(\phi) = 1 + \xi (\phi/M_{\rm pl})^n$ with $n$ being the power law index is adopted. In particular,  $A(\phi) = 1 + \xi (\phi /M_{\rm pl})^2$ is assumed for the Higgs inflation case. In some works, a cosine form  $A(\phi) = 1 + \xi ( 1 + \cos (\phi/f))^n$ with $f$ being a model parameter is assumed  for natural inflation as the Jordan frame potential \cite{Ferreira:2018nav}. Regarding the Jordan frame potential $V_J$, one can take an arbitrary form, at least phenomenologically,  however, in most works of non-minimal inflation, rather limited ones have been assumed such as a power-law type $V_J \propto \phi^p$ with $p$ being a positive integer, a cosine type  $V_J \propto 1+ \cos (\phi/f)$ and so on.  

In this paper, we discuss the the predictions for inflationary observables such as $n_s$ and $r$ in a general setting although we also take some explicit forms for $V_J$ and ${\cal F}$ as examples to present how the general expressions actually work.  To make a systematic analysis, we assume that the Jordan frame potential satisfies  $d V_J / d\phi > 0$ in which the inflaton for the minimally coupled case moves from a (positively) large field value to a (positively) smaller one towards the end of inflation for the range of $\phi$ relevant to the inflationary dynamics. We also assume that the inflation ends by the violation of slow-roll when $\epsilon = 1$ is satisfied both in minimally and non-minimally coupled cases\footnote{
Even when the inflation does not end by the slow-roll violation in a minimally coupled case,  the end of inflation can be invoked by the slow-roll violation due to the presence of a non-minimal coupling \cite{Takahashi:2020car}. However, in this paper, we do not consider such kind of models. 
}. 

It is important to notice that when the non-minimal coupling exists, the Einstein frame potential $V_E$ does not necessarily satisfy $ d V_E / d\chi > 0$ even if  the original Jordan frame potential fulfills $d V_J / d\phi > 0$ for the corresponding range of $\phi$ (or $\chi$),  which motivates us to classify models using the forms of $A(\phi)$ (or ${\cal F}(\phi)$) and $V_J(\phi)$.  We discuss this issue in the next section.

\subsection{Classification of models \label{sec:classification}}

The derivative of the Einstein frame potential~\eqref{eq:V_E}  with respect to $\chi$ is written as
\begin{equation}
\frac{d V_E}{d\chi} 
= 
\frac{d\phi}{d\chi} \left( \frac{V_J}{A^2} \right) \left(  \frac{V_J'}{V_J} - 2 \frac{A'}{A}  \right)   \,.
\end{equation}
When $\xi$ is sufficiently large, this  can be expanded as 
\begin{equation}
\frac{d V_E}{d\chi} 
=
\frac{d\phi}{d\chi} \left( \frac{V_J}{{\cal F}^2} \right) \left(  \frac{V_J'}{V_J} - 2\frac{{\cal F}'}{{\cal F}}  \right) \frac{1}{\xi^2} + {\cal O}\left( \frac{1}{\xi^3} \right) \,.
\end{equation}
Notice that even if we assume that $V_J ' >0$ holds, the derivative of the Einstein frame potential can be positive, zero or negative for sufficiently large $\xi$ and $\chi$, 
depending on the sign of  $V_J' / V_J -  2 {\cal F}'/{\cal F}$, which is determined by the functional forms of $V_J$ and ${\cal F}$. This motivates us to classify models into three categories as follows:
\begin{eqnarray}
\label{eq:attractor_type}
 \frac{V_J'}{V_J} = 2 \frac{{\cal F}'}{{\cal F}}     &\qquad & ({\rm Attractor~type}), \\ [8pt]
\label{eq:VJ_dominant}
  \frac{V_J'}{V_J} > 2 \frac{{\cal F}'}{{\cal F}}     &\qquad & (V_J{\rm -dominant~type}), \\ [8pt]
\label{eq:F_dominant}
  \frac{V_J'}{V_J} < 2 \frac{{\cal F}'}{{\cal F}}     &\qquad & ({\cal F}{\rm -dominant~type}), 
\end{eqnarray} 
which respectively correspond to $dV_E/ d\chi =0, dV_E /d\chi >0$ and $dV_E /d\chi  < 0$ at large $\xi$ and $\chi$ limit. 

For later convenience, here we introduce the parametrization for the relation between $V_J$ and ${\cal F}$ as 
\begin{equation}
\label{eq:VJ_F_alpha}
\alpha \frac{V_J'}{V_J}  =  2 \frac{\cal F'}{\cal F}  \,,
\end{equation}
where we assume $\alpha$ to be constant in the following argument. Although in general $\alpha$ can depend on $\phi$, most models studied in the literature so far correspond to the case that $\alpha$ is constant. Moreover during the slow-roll phase and around the time when the scales of interest exit the horizon, $\alpha$ would not change much even if it depends on $\phi$. Thus the assumption that $\alpha$ is constant would give a good approximation when we discuss inflationary observables.  By using $\alpha$, each class given above can be specified  as follows:
\begin{eqnarray}
\alpha=0    &\qquad & ({\rm Attractor~type}), \\ [8pt]
0 < \alpha < 1   &\qquad & (V_J{\rm -dominant~type}), \\ [8pt]
1< \alpha   &\qquad & ({\cal F}{\rm -dominant~type}). 
\end{eqnarray}

For large values of $\xi$ and  $\chi$, the Einstein frame potential $V_E$ for  the attractor type exhibits a very flat shape.  In the $V_J$-dominant type, $V_E$ is a monotonically increasing function  at least for the range of $\phi$ relevant to the inflationary dynamics.  In the ${\cal F}$-dominant type, $V_E$ has an extremum at some value of $\chi$ (or $\phi$), which we denote as $\chi_{\rm ex}$ (or $\phi_{\rm ex}$),  for a sufficiently large value of $\xi$. In the following discussion,  we consider the range of $\chi \, (\phi)$ as $\chi < \chi_{\rm ex} ~(\phi < \phi_{\rm ex})$ in order that $\chi  \,(\phi)$ does not run away towards a large value, which is usually assumed.  Actually as $\xi$ gets larger, the value of $\chi$ should be taken to be very close to $\chi_{\rm ex}$ to realize a sufficient number of $e$-folds ($N = 50-60$), which gives a distinctive feature for ${\cal F}$-dominant case. Such kinds of features will be discussed in Sec.~\ref{sec:predictions}.

\subsection{Example models \label{sec:example_models}}

Although the aim of this paper is to give predictions for the inflationary observables in a general setting, i.e., regardless of the forms of the Jordan frame potential $V_J (\phi)$ and the functional form for non-minimal coupling ${\cal F} (\phi)$ as much as possible, some explicit examples are useful to understand the behavior and applicability of our approach. In this paper, we consider three examples which are explained in the following.

\subsubsection{Chaotic inflation with power-law ${\cal F}(\phi)$}

One of explicit models we consider is the chaotic inflation in the Jordan frame with a power-law form ${\cal F}$ in which the Jordan frame potential and the function ${\cal F}$ are respectively given by 
\begin{eqnarray}
\label{eq:VJ_chaotic}
V_J (\phi)  &=&  V_0 \left( \frac{\phi}{\Mpl} \right)^p \,,  \\ [8pt]
\label{eq:F_chaotic}
{\cal F} (\phi)  &=&   \left( \frac{\phi}{\Mpl} \right)^n \,, 
\end{eqnarray}
where $V_0$ is a parameter representing the energy scale for the potential,  and $p$ and $n$ are the power-law indices for $V_J$ and ${\cal F}$, respectively.  As mentioned in the introduction, the minimally coupled version of chaotic inflation for any $p$  has been excluded by  the constraints on $n_s$ and $r$ from Planck+BAO+BICEP/Keck 2018 \cite{BICEP:2021xfz}. The Jordan frame slow-roll parameters are given by
\begin{equation}
\label{eq:SR_Jordan_chaotic}
\epsilon_J = \frac{p^2}{2}\left( \frac{\Mpl}{\phi} \right)^2, \qquad \eta_J = p(p-1)\left( \frac{\Mpl}{\phi} \right)^2 \,.
\end{equation}
From these expressions, one can find that $\epsilon_J$ and $\eta_J$  are related as
\begin{equation}
\eta_J = \frac{2(p-1)}{p} \epsilon_J \,,
\end{equation}
which is used to derive some expressions in the following section.

Once $p$ and $n$ are given,  the $\alpha$ parameter is written as 
\begin{equation}
\alpha = \frac{2n}{p} \,,
\end{equation}
and then this model can be classified into three types as follows:
\begin{eqnarray}
p = 2 n && ({\rm attractor~type})  \,, \\  [8pt]
p > 2 n && (V_J{\rm -dominant~type})  \,, \\  [8pt]
p < 2 n &&({\cal F}{\rm -dominant~type})    \,.
\end{eqnarray}
The Higgs inflation  corresponds to the case of $p=4$ and $n=2$, which is classified as the attractor type.

\begin{figure}[t]
\begin{center}
\includegraphics[width=5.5cm]{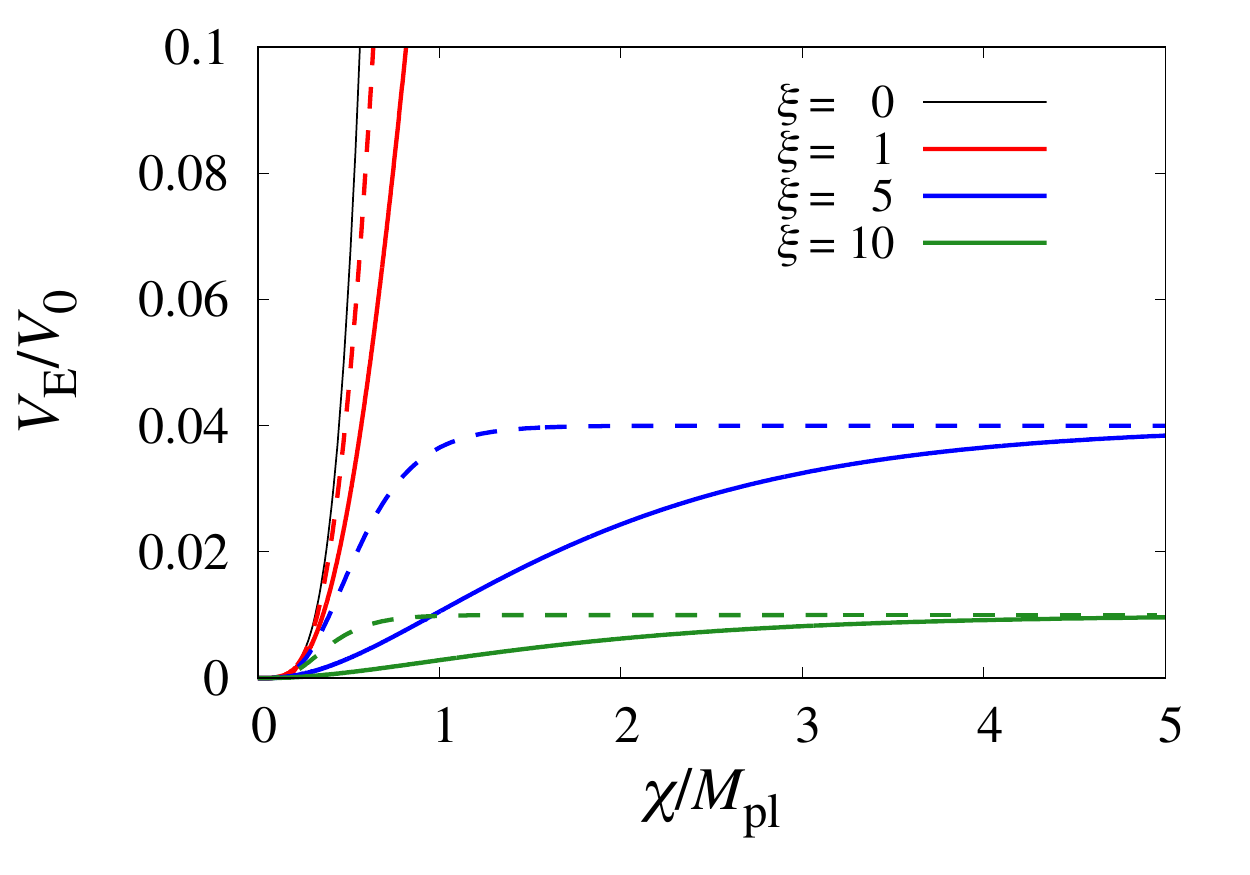}\includegraphics[width=5.5cm]{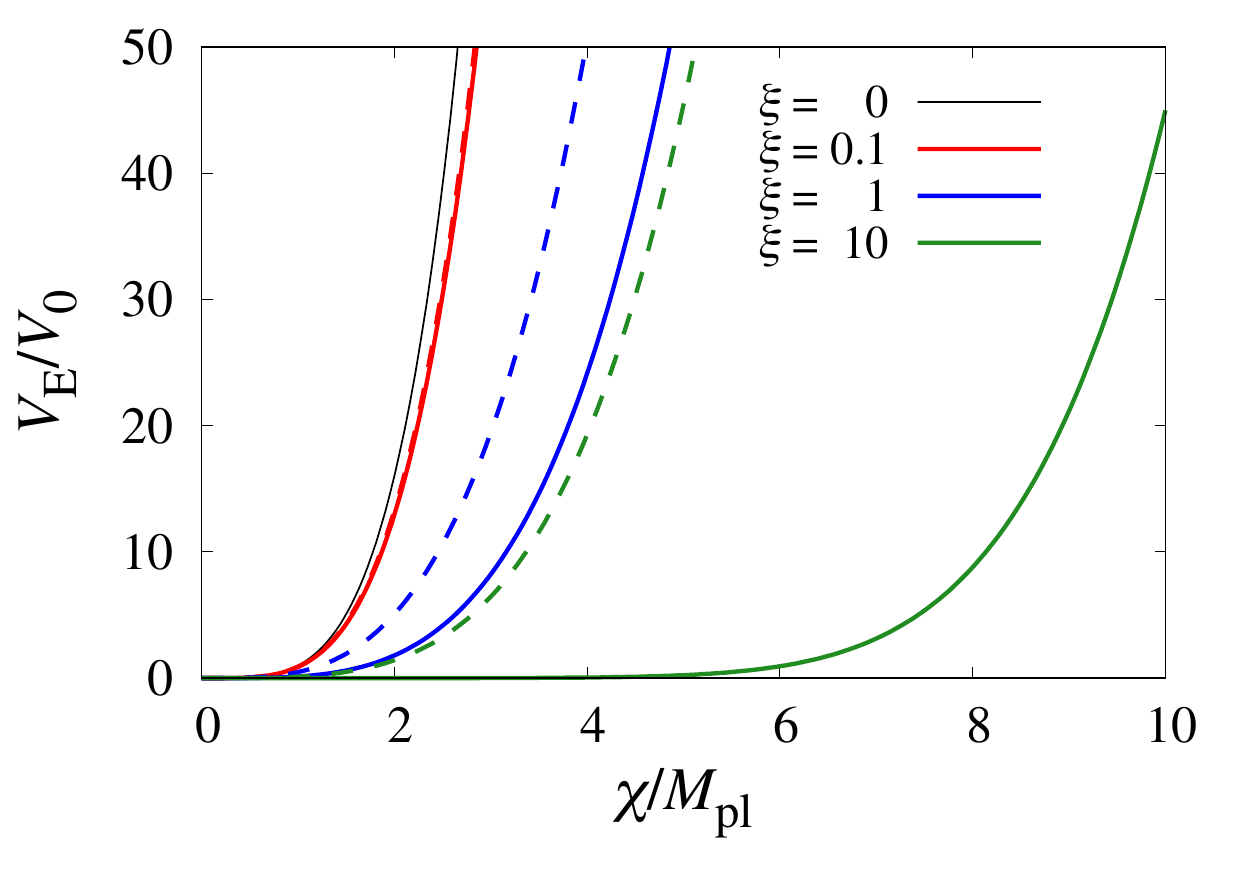}\includegraphics[width=5.5cm]{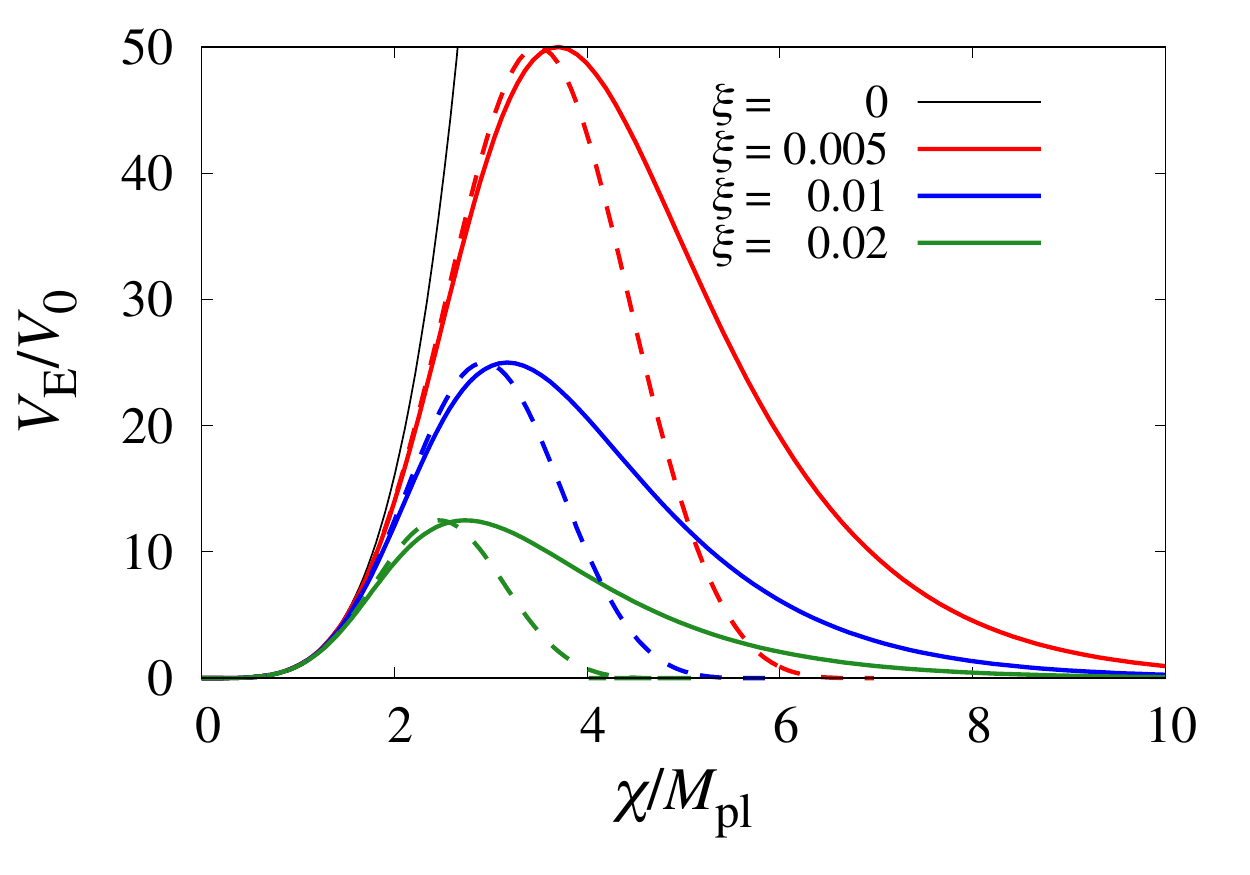}\\
\end{center}
\caption{\label{fig:VE_chaotic} Einstein frame potential for  $V_J =V_0 (\phi / M_{\rm pl})^p$ and ${\cal F} = (\phi / M_{\rm pl})^n$ with $(p,n)=(4,2)$ (left), $(p,n)=(4,1)$ (middle) and $(p,n)=(4,4)$ (right), which correspond to the attractor, $V_J$-dominant and ${\cal F}$-dominant types, respectively.   We take several values of $\xi$ for each case which are shown in the figure. Solid and dashed lines represent the metric and Palatini formulation cases, respectively.}
\end{figure}

In Fig.~\ref{fig:VE_chaotic}, we show the Einstein frame potential for some combinations of $p$ and $n$. The case with  $(p,n)=(4,2)$ (left panel) corresponds to the attractor-type.  As seen from the figure, this type gives a flat shape in the large $\chi$ region as $\xi$ increases.  The case with  $(p,n)=(4,1)$ (middle panel) exhibits the $V_J$-dominant type, in which the potential is just a monotonically increasing function as the Jordan frame one.  The case with  $(p,n)=(4,4)$ (right panel) is an example of ${\cal F}$-dominant type in which  an extremum appears at some $\chi$   as seen from the figure. Both  the metric and Palatini formulation cases are shown in the figure.

\subsubsection{Natural inflation with cosine-type ${\cal F}(\phi)$}

Another model we consider  is the natural inflation in which the Jordan frame potential and a cosine-type ${\cal F}$ are assumed as 
\begin{eqnarray}
V_J (\phi)  &=&  V_0 \left[  1 - \cos \left( \frac{\phi}{f} \right) \right] \,,  \\ [8pt]
{\cal F} (\phi)  &=&   \left[  1 - \cos \left( \frac{\phi}{f} \right) \right]^n \,, 
\end{eqnarray}
where $f$ is a model parameter and $n$ is the power-law index for ${\cal F}$. The case with $n=1$ has been discussed in \cite{Ferreira:2018nav}. In this paper, we consider the model with an arbitrary value for $n$.   As mentioned in the introduction, the minimally coupled counterpart has been excluded by the result from Planck+BAO+BICEP/Keck 2018 \cite{BICEP:2021xfz}.  The Jordan frame slow-roll parameters are given by 
\begin{equation}
\label{eq:eps_J_natural}
\epsilon_J = \frac{1}{2}\left( \frac{\Mpl}{f} \right)^2 \cot^2 \left( \frac{\phi}{2f} \right), 
\qquad 
\eta_J = \left( \frac{\Mpl}{f} \right)^2 \frac{1 - 2 \sin^2  \left( \frac{\phi}{2f} \right)}{2 \sin^2  \left( \frac{\phi}{2f} \right)} \,.
\end{equation}
Actually $\epsilon_J$ and $\eta_J$ are related as 
\begin{equation}
\label{eq:eta_eps_natural}
\eta_J = \epsilon_J - \frac12 \left( \frac{\Mpl}{f}\right)^2 \,,
\end{equation}
which is useful to simplify some expressions in later discussion. 

The $\alpha$ parameter in this model is given by 
\begin{equation}
\alpha = 2n \,,
\end{equation}
and then three categories in the classification given in Sec.~\ref{sec:classification} are specified as
\begin{eqnarray}
\label{eq:n_attaractor}
n=\frac12  && ({\rm attractor~type})  \,, \\  [8pt]
n < \frac12 && (V_J{\rm -dominant~type})  \,, \\  [8pt]
\label{eq:n_Fdom}
n > \frac12 &&({\cal F}{\rm -dominant~type})    \,.
\end{eqnarray}

\begin{figure}[t]
\begin{center}
\includegraphics[width=5.5cm]{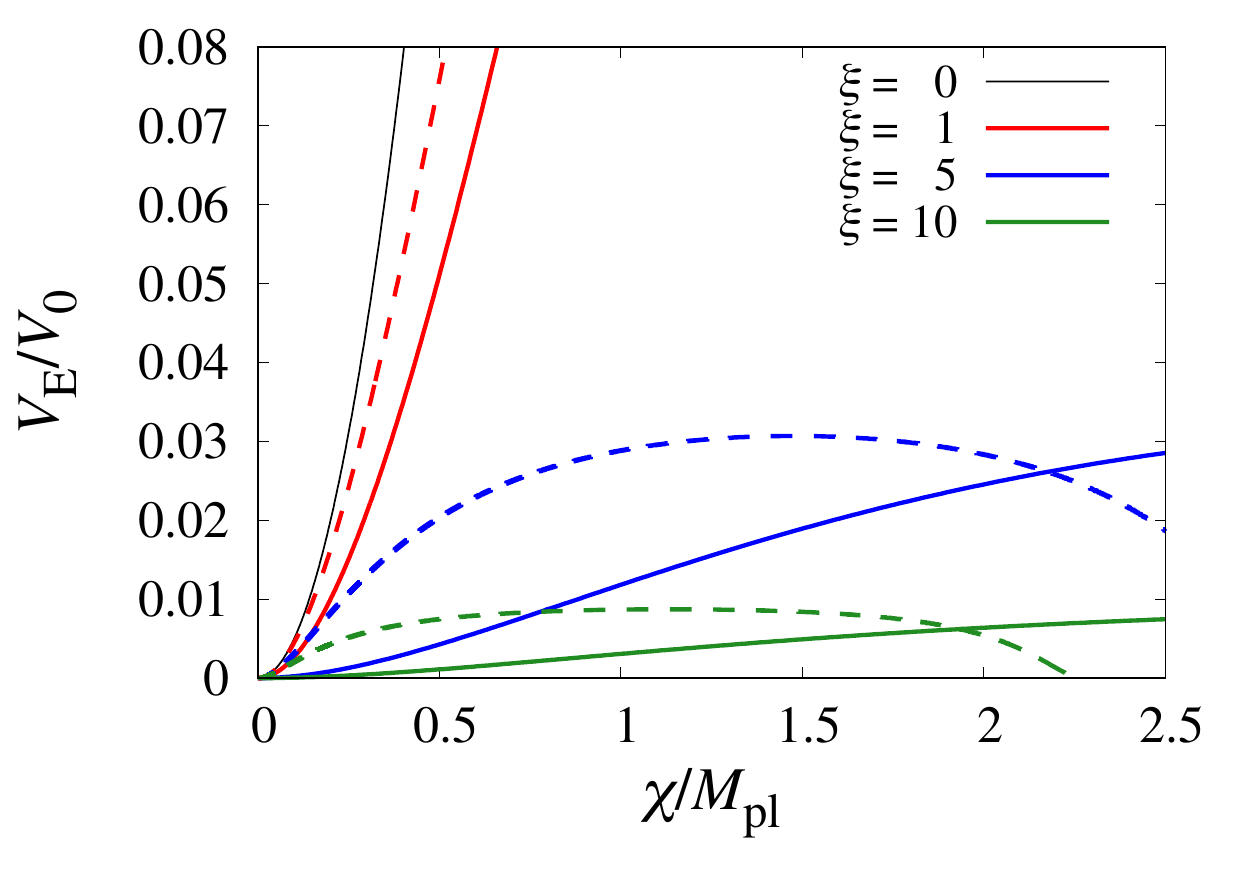}\includegraphics[width=5.5cm]{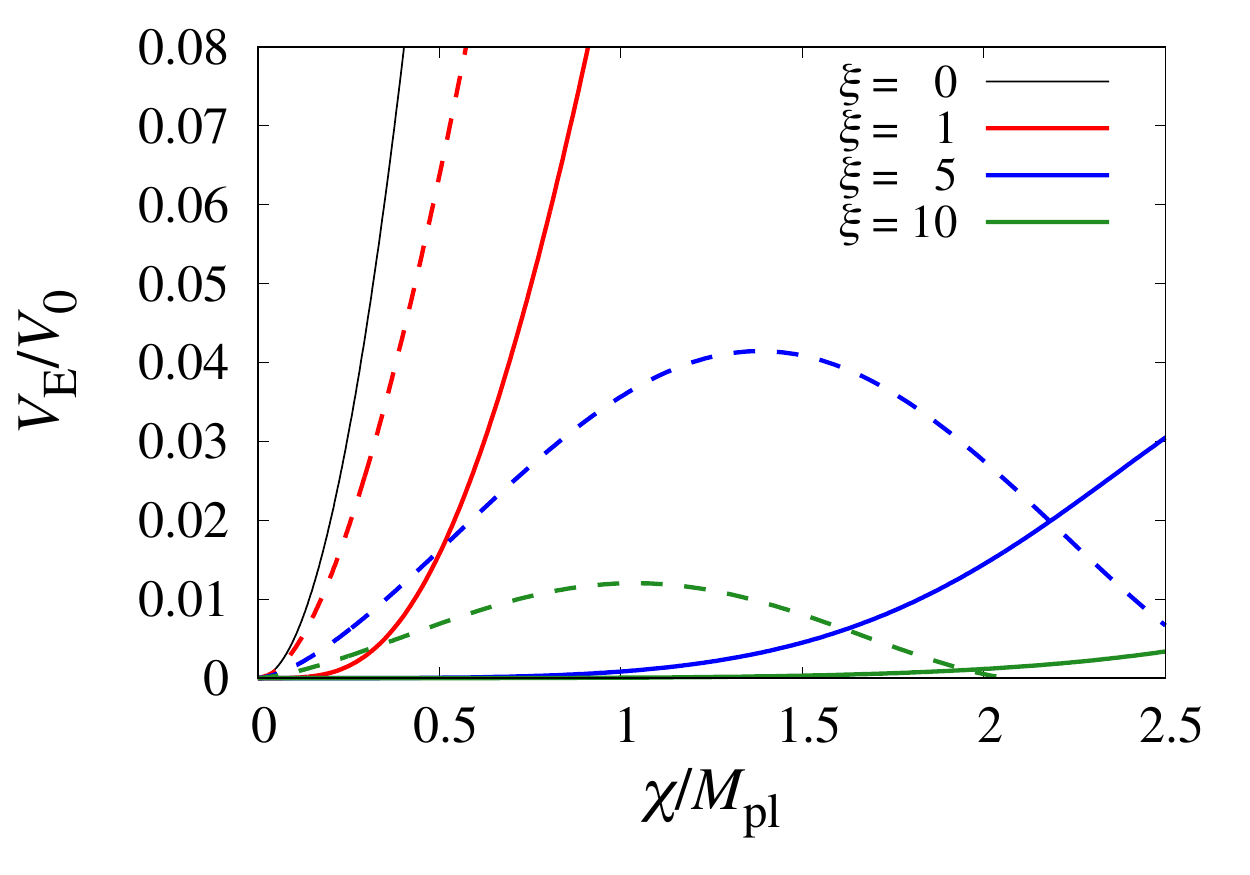}\includegraphics[width=5.5cm]{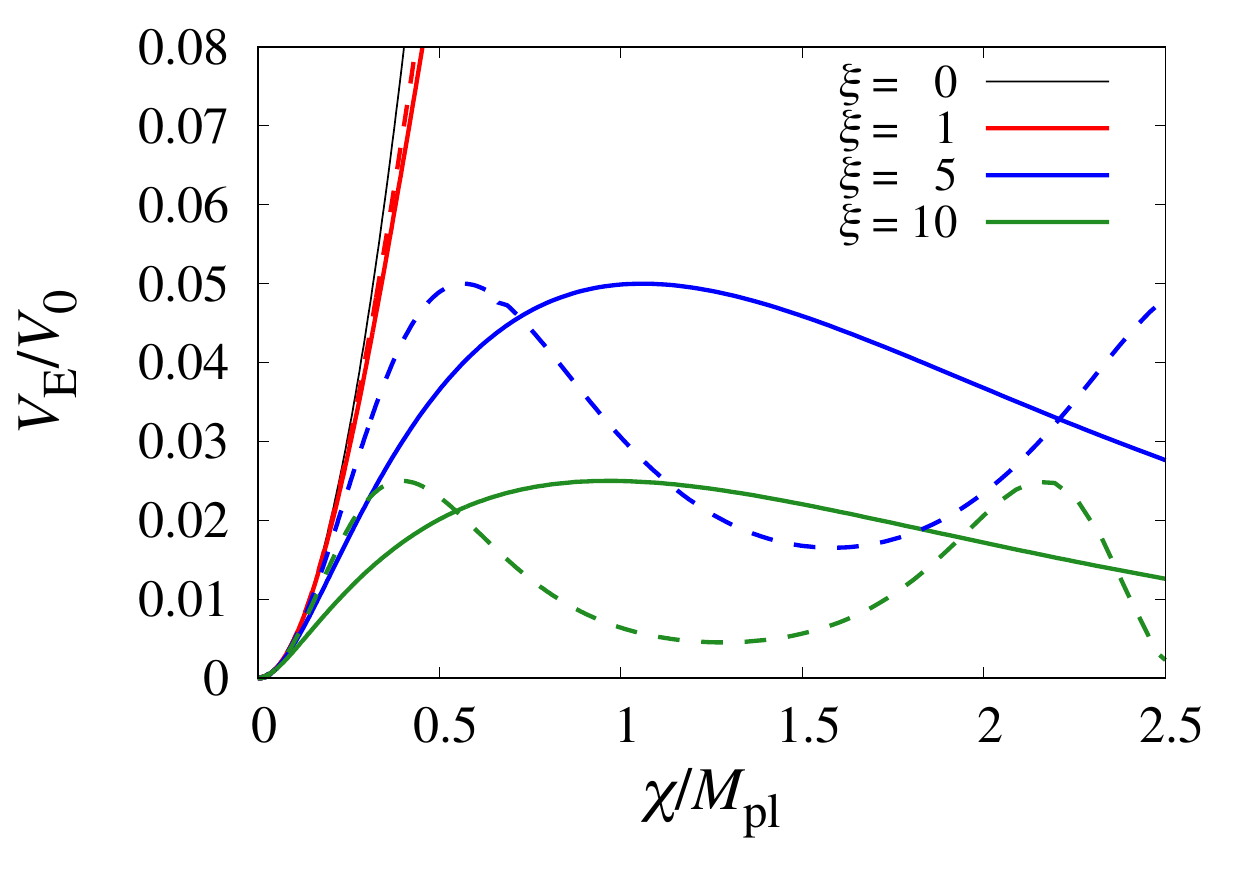}\\
\end{center}
\caption{\label{fig:VE_natural} Einstein frame potential for  $V_J =V_0 \left[ 1 - \cos (\phi /f) \right] $ and ${\cal F} = \left[ 1 - \cos (\phi /f \right)]^n$ with $n=1/2$ (attractor type: left), $n=1/4$ ($V_J$-dominant type:  middle) and $n=1$ (${\cal F}$-dominant type:  right) are  shown.  We take $f = \Mpl$ in this figure. Several values of $\xi$ are taken for each case whose values are written in the figure.  Solid and dashed lines represent the metric and Palatini formulation cases, respectively.}
\end{figure}

In Fig.~\ref{fig:VE_natural}, the Einstein frame potential in this model is shown for $n=1/2, 1/4$ and $1$, which respectively correspond to the attractor,  $V_J$-dominant and ${\cal F}$-dominant types,  for several values of $\xi$. Since the Jordan frame potential in this model already has an extremum at $\phi_{\rm ex} = \pi f$, the Einstein frame potential also has an extremum for every category, however, each class gives distinctive features for $n_s$ and $r$, which will be discussed in the next section.

\subsubsection{Loop inflation with log-type ${\cal F}(\phi)$}
The other model considered in this paper is the loop inflation\footnote{
This model is referred as ``loop inflation" in \cite{Martin:2013tda} and ``Spontaneously broken SUSY" model in \cite{Planck:2018jri}.
} where the Jordan frame potential and a log-type functional form for $\cal{F}$ are assumed as
\begin{eqnarray}
V_J(\phi) &=& V_0\left[ 
1 + a_h \log\left(\frac{\phi}{\Mpl}\right)
\right]\,, \\ [8pt]
\cal{F} &=& \left[
1 + a_h \log\left(\frac{\phi}{\Mpl}\right) 
\right]^n \,,
\end{eqnarray}
where $a_h$ is a model parameter and $n$ is the power-law index for $\cal{F}$. We assume that  $a_h > 0$ in order that the inflaton moves from a large positive value to a small positive one.  The $\alpha$ parameter in this model is given as $ \alpha = 2n$, which is the same as the one in the natural inflation model with the cosine-type ${\cal F}$ and the classification is also given in the same way as Eqs.~\eqref{eq:n_attaractor}--\eqref{eq:n_Fdom}.

To have $V_J (\phi) >0$, the field value should satisfy  $\phi/\Mpl > e^{-1/a_h}$. The minimally coupled counterpart of this model for  $a_h \lesssim 0.5$ has been excluded by the constraints on $n_s$ from Planck \cite{Planck:2018jri} since it gives a bluer spectral index than the observational limit.  The functional form for ${\cal F}$, i.e., a log-type ${\cal F}$,  is chosen to keep $\alpha$ constant for the whole range of $\chi (\phi)$ relevant to the inflationary dynamics.  Although this choice is just a phenomenological one, the trend for the predictions of the inflationary observables would be similar as far as the value of $\alpha$ is almost the same. The Jordan frame slow-roll parameters in this model are given by
\begin{align}
\label{eq:eps_J_loop}
\epsilon_J = \frac{1}{2}\left(\frac{\Mpl}{\phi}\right)^2 \left[
\frac{a_h}{1 + a_h\log\left(\frac{\phi}{\Mpl}\right)}
\right]^2\,, \qquad
\eta_J = -\left(\frac{\Mpl}{\phi}\right)^2 \frac{a_h}{1 + a_h \log\left(\frac{\phi}{\Mpl}\right)} \,.
\end{align}
From these expressions, $\epsilon_J$ and $\eta_J$ can be related as 
\begin{equation}
\label{eq:Jordan_eps_eta_loop}
\eta_J = -2 \epsilon_J \frac{a_h}{1 + a_h \log (\phi / M_{\rm pl} ) } \,.
\end{equation}
This relation is useful to derive some expressions in the next section.

\begin{figure}[t]
\begin{center}
\includegraphics[width=5.5cm]{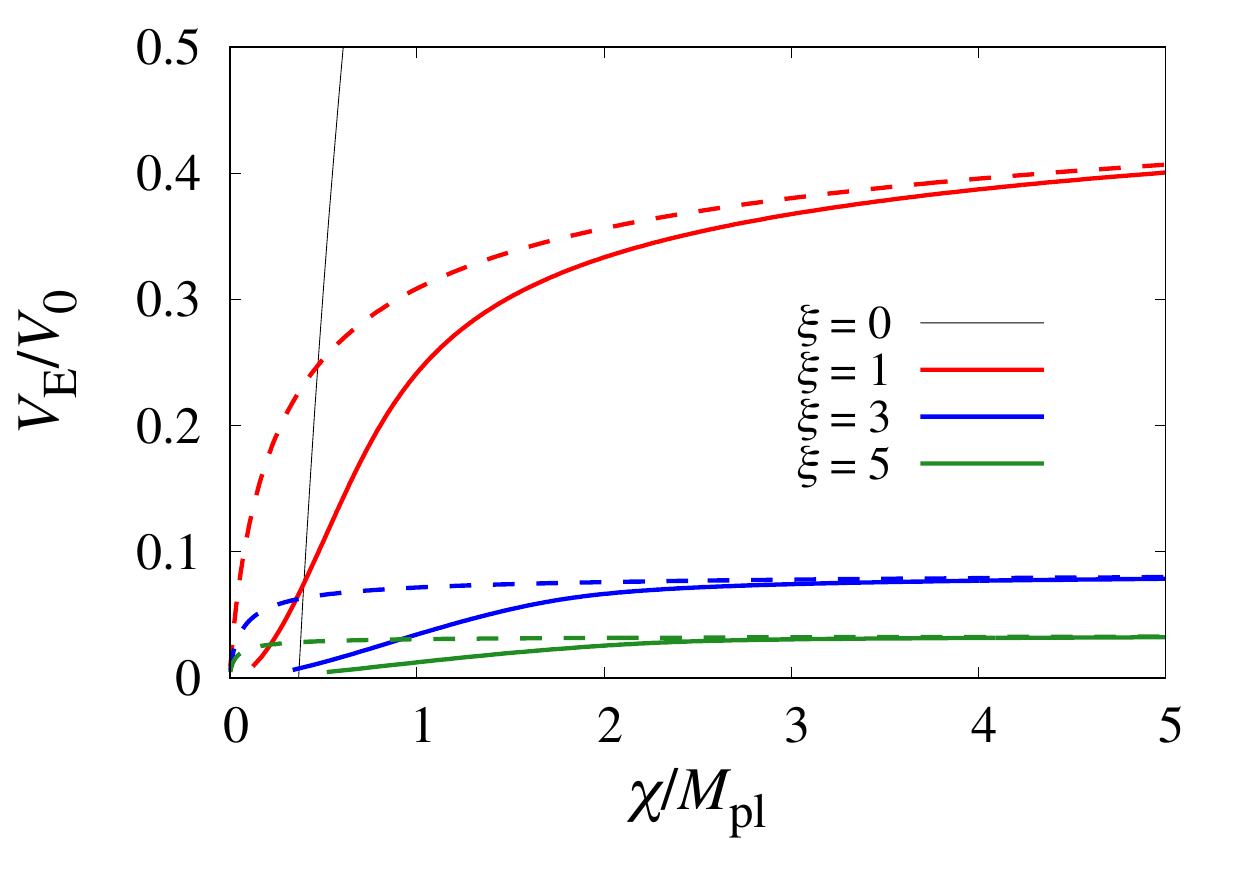}\includegraphics[width=5.5cm]{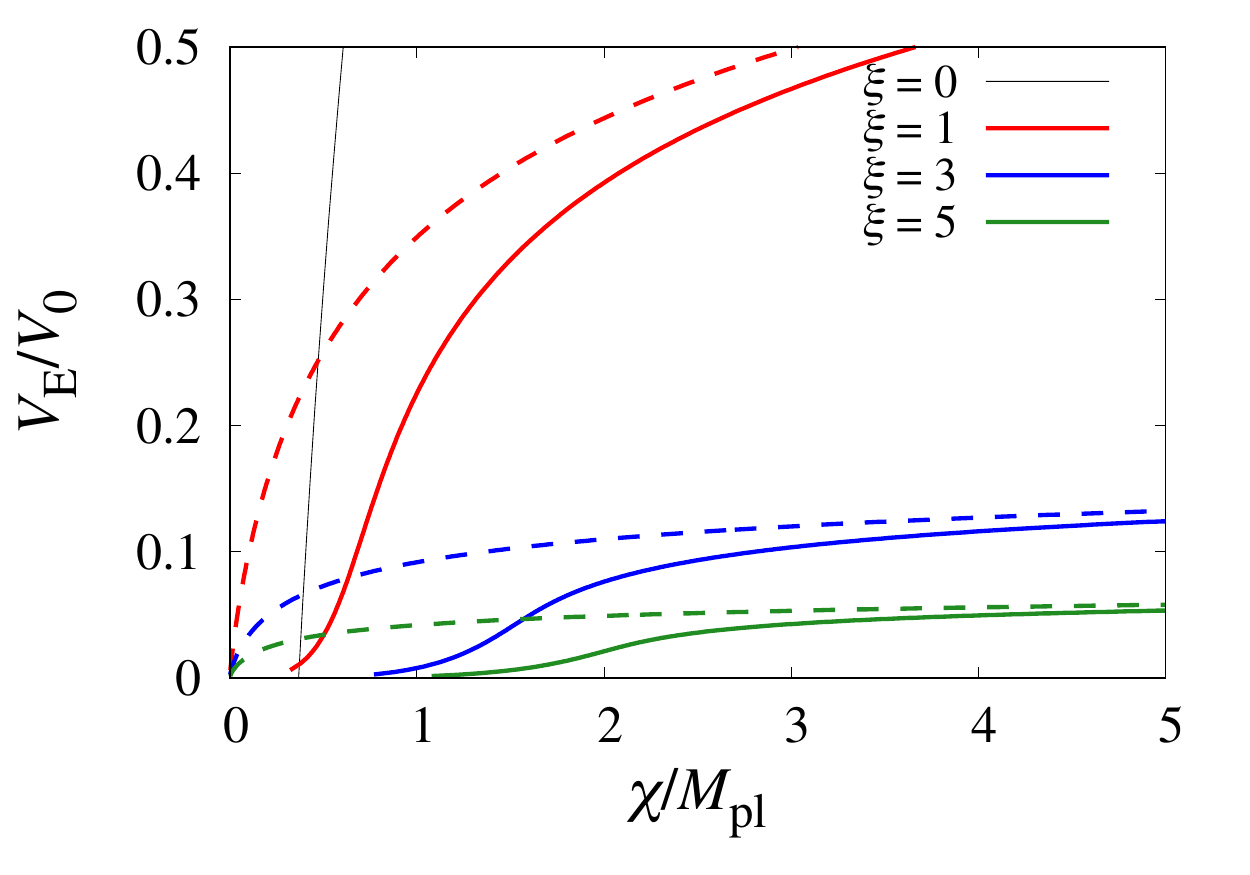}\includegraphics[width=5.5cm]{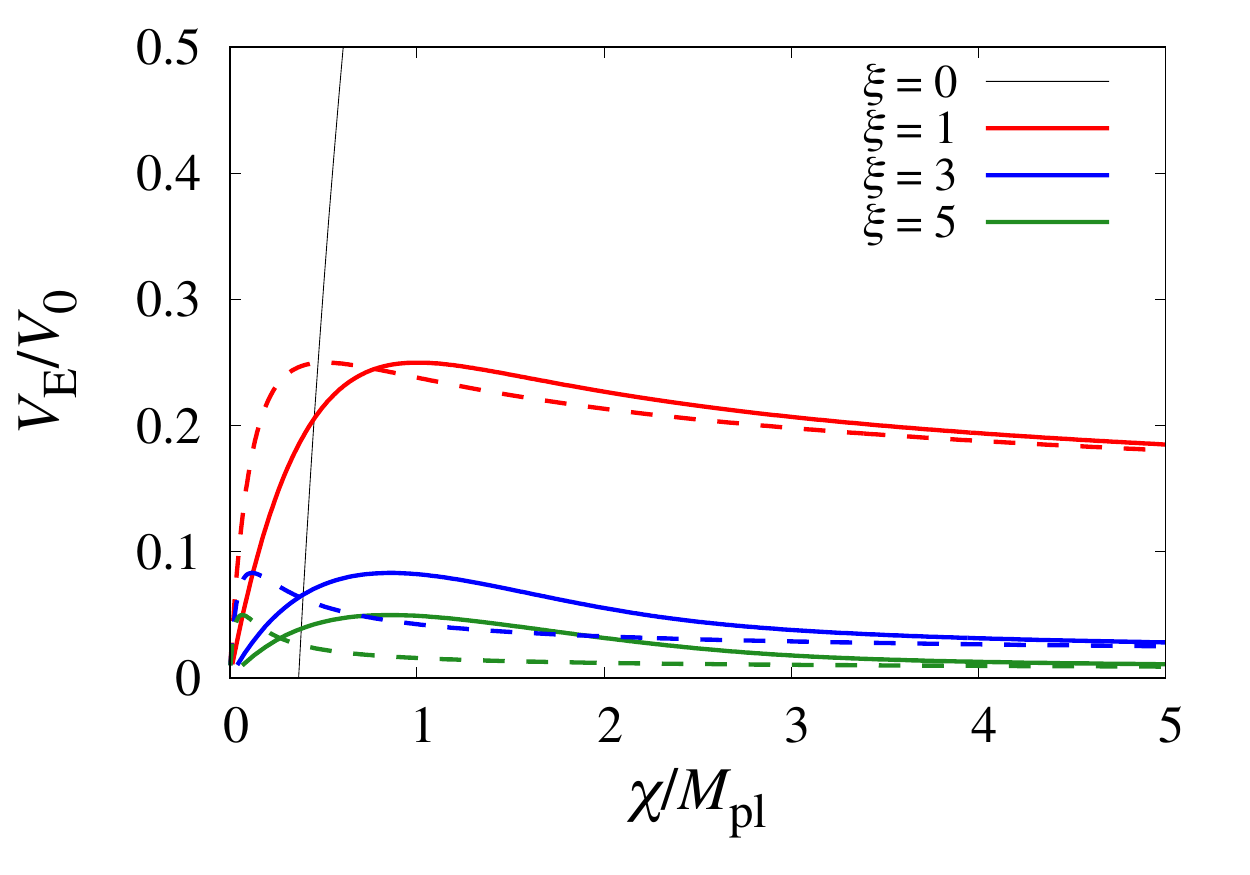}\\
\end{center}
\caption{\label{fig:VE_loop} Einstein frame potential for  $V_J =V_0\left[ 1 + a_h \log\left( \phi / \Mpl \right)\right]$ and ${\cal F} = \left[ 1 + a_h \log\left( \phi / \Mpl \right) \right]^n$ with $n=1/2$ (attractor type: left), $n=1/4$ ($V_J$-dominant type:  middle) and $n=1$ (${\cal F}$-dominant type:  right) are  shown.  We take several values of $\xi$ for each case which are shown in the figure.  Solid and dashed lines represent the metric and Palatini formulation cases, respectively.}
\end{figure}

In Fig.~\ref{fig:VE_loop}, the Einstein frame potential for this model is shown for $n=1/2, 1/4$ and $1$, which correspond to the attractor, $V_J$-dominant and ${\cal F}$-dominant types, respectively.  Since the Jordan frame potential is a monotonically increasing function, the shape of the Einstein frame potential is basically the same as the ones for the chaotic inflation with power-law type ${\cal F}$ discussed above.

\bigskip\bigskip\bigskip
Having described the framework and some assumptions adopted in the analysis, as well as some explicit models to be considered as an example,  then in the next section, we discuss general predictions and some explicit examples  for each type.

\section{Predictions for the inflationary observables \label{sec:predictions}}

Now in this section, we investigate predictions for inflationary observables, first without specifying the Jordan frame potential $V_J$ and the non-minimal coupling function ${\cal F}$, followed by the arguments using some explicit models presented in the previous section for illustrations.  Models for each type introduced in Eqs.~\eqref{eq:attractor_type}--\eqref{eq:F_dominant}  give peculiar predictions for the inflationary observables such as the spectral index and the tensor-to-scalar ratio, whose features are discussed in order below.

\subsection{Attractor type}

The attractor type is specified by the relation
\begin{equation}
\label{eq:attractor_VJ_F2}
\frac{V_J'}{V_J} = 2 \frac{{\cal F}'}{{\cal F}}  \,,
\end{equation}
from which one can relate $V_J$ and ${\cal F}$ as 
\begin{equation}
\label{eq:VJ_F2}
V_J = C {\cal F}^2 \,,
\end{equation}
with $C$ being a constant. In this case, the slow-roll parameters in the large $\xi$ limit for the metric case can be expanded, at the leading order,  as
\begin{eqnarray}
\label{eq:slow_roll_Einstein_attractor_eps}
&& \epsilon = \frac{4}{3 {\cal F}^2 \xi^2}   \,, \\ [8pt] 
\label{eq:slow_roll_Einstein_attractor_eta}
&& \eta = -  \frac{4}{3 {\cal F} \xi} \,. 
\end{eqnarray}
The number of $e$-folds for the large $\xi$ limit is calculated as 
\begin{equation}
\label{eq:N_large_xi_metric}
N = \frac1{M_{\rm pl}^2} \int_{\phi_{\rm end}}^{\phi_\ast} \frac34 {\cal F}' \xi  \,\,d\phi
\simeq 
\frac34 \xi {\cal F}(\phi_\ast) \,,
\end{equation}
where we have neglected the contribution from $\phi_{\rm end}$. 

By using Eqs.~\eqref{eq:slow_roll_Einstein_attractor_eps}, \eqref{eq:slow_roll_Einstein_attractor_eta} and \eqref{eq:N_large_xi_metric}, one can express $\epsilon$ and $\eta$ as 
\begin{eqnarray}
&& 
\epsilon = \frac{3}{4N^2} \,,  
\qquad
 \eta = -\frac1N \,,
\end{eqnarray}
from which the spectral index and the tensor-to-scalar ratio are given as 
\begin{eqnarray}
&& 
\label{eq:attractor_ns}
n_s -1  =  -6 \cdot \frac{3}{4N^2}  - 2 \frac1N  \simeq - \frac2N \,,  \\ [8pt] 
&&
\label{eq:attractor_r}
r = 16 \cdot \frac{3}{4N^2} = \frac{12}{N^2}  \,.
\end{eqnarray}
One can see that the predictions for $n_s$ and $r$ do not depend on $\xi$ in the large $\xi$ limit and exhibit the attractor nature of this type.   It should be noted that when the number of $e$-folds is $N=50-60$ as commonly assumed, the predictions for $n_s$ and $r$ are given by
\begin{equation}
n_s \simeq 0.960 - 0.967 \,, 
\qquad 
r \simeq 0.0048 - 0.0033 \,,
\end{equation}
which are well within the current observational bound \cite{BICEP:2021xfz}. An attractor behavior of this type has  been discussed in \cite{Park:2008hz,Kallosh:2013tua} (see also \cite{Kallosh:2021mnu,Cheong:2021kyc} for the discussion after the result of BICEP/Keck 2018 \cite{BICEP:2021xfz}).

On the other hand, the Palatini case gives the following predictions for the slow-roll parameters in the large $\xi$ limit at leading order as
\begin{eqnarray}
\label{eq:eps_large_xi_palatini}
&& \epsilon =  
 \frac{2 ({\cal F'})^2 M_{\rm pl}^2}{{\cal F}^3 \xi } \,, \\ [8pt]  
&& \eta =  
M_{\rm pl}^2 \left(  - 3  \left( \frac{\cal F'}{\cal F} \right)^2 + 2  \frac{\cal F^{\prime\prime}}{\cal F} \right)
+  \frac{5 ({\cal F'})^2 M_{\rm pl}^2}{{\cal F}^3 \xi }   \,.
\end{eqnarray}
The number of $e$-folds for the potential~\eqref{eq:VJ_F2} is given by 
\begin{equation}
N = \frac1{M_{\rm pl}^2} \int_{\phi_{\rm end}}^{\phi_\ast} \frac{\cal F}{2\cal F'}   \,\,d\phi \,,
\end{equation}
from which one can see that $N$ in the large $\xi$ limit for the Palatini case does not depend on $\xi$. 

Actually when one uses  the relation~\eqref{eq:attractor_VJ_F2}, $N$ can be rewritten, by using $V_J$,  as 
\begin{equation}
N = \frac1{M_{\rm pl}^2} \int_{\phi_{\rm end}}^{\phi_\ast} \frac{V_J}{V_J'}   \,\,d\phi \,.
\end{equation}
As shown in Eq.~\eqref{eq:eps_large_xi_palatini}, when $\xi$ is very large, $\epsilon$ decreases much. On the other hand, with the relation $V_J = C {\cal F}^2$, $\eta$ can be rewritten as 
\begin{eqnarray}
&& \eta =  -\frac52 \epsilon_J  + \eta_J \,.
\end{eqnarray}
Since $\epsilon$ can be neglected in the large $\xi$ limit as mentioned above, the spectral index is given by
\begin{eqnarray}
&& 
\label{eq:ns_attractor_palatini}
n_s -1  \simeq 2 \eta = -5 \epsilon_J  + 2\eta_J \,.
\end{eqnarray}
From Eq.~\eqref{eq:eps_large_xi_palatini} and by using the relation for the attractor type~\eqref{eq:attractor_type},  the the tensor-to-scalar ratio can be written, at the leading order in $1/\xi$, as 
\begin{equation}
\label{eq:attractor_r_palatini}
r = \frac{16 \epsilon_J}{{\cal  F} \xi} \,,
\end{equation}
from which one can see that $r$ is very suppressed when $\xi$ becomes large.

\begin{figure}[htbp]
\begin{center}
\includegraphics[width=16.5cm]{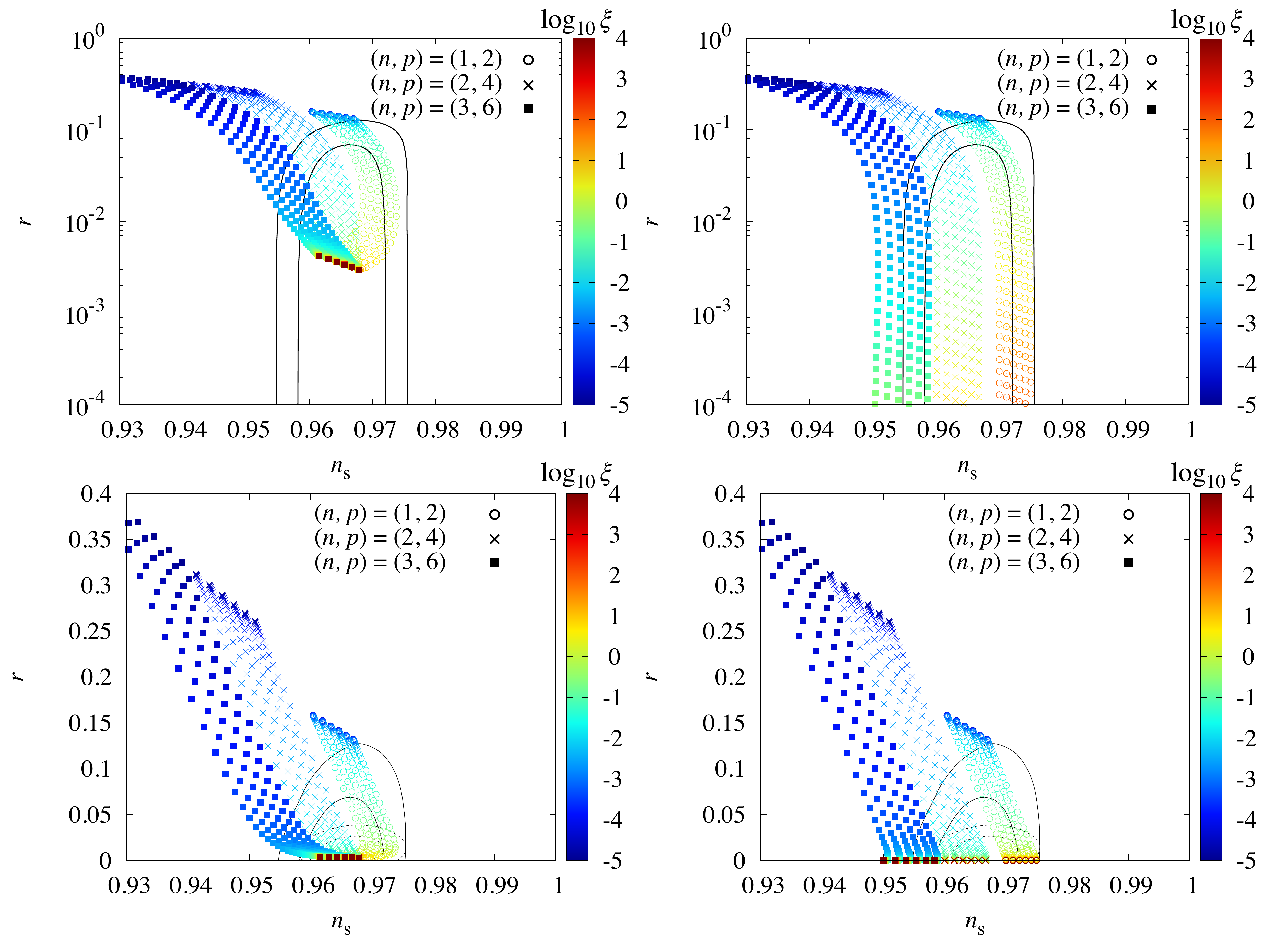}
\end{center}
\caption{\label{fig:attractor_chaotic}  Predictions of $n_s$ and $r$ in chaotic inflation with power-law ${\cal F}$ for several values of $p$ and $n$ giving the attractor type whose values are shown in the figure. Cases for the metric formulation (left column panels) and Palatini formulation (right column panels) are shown.  Top and bottom panels for each column are the same except that $r$ is shown by logarithmic and linear scales, respectively. The value of $\xi$ is represented by color as shown in the color bar in the legend. The number of $e$-folds is assumed as $N=50-60$. 1$\sigma$ and 2$\sigma$ constraints from Planck \cite{Planck:2018jri}  and Planck+BAO+BICEP/Keck~2018 \cite{BICEP:2021xfz} are also depicted with solid and dashed lines, respectively.  When $r$ is shown in a log-scale (top panels), we only show the Planck constraint.}
\end{figure}

\begin{figure}[ht]
\begin{center}
\includegraphics[width=16.5cm]{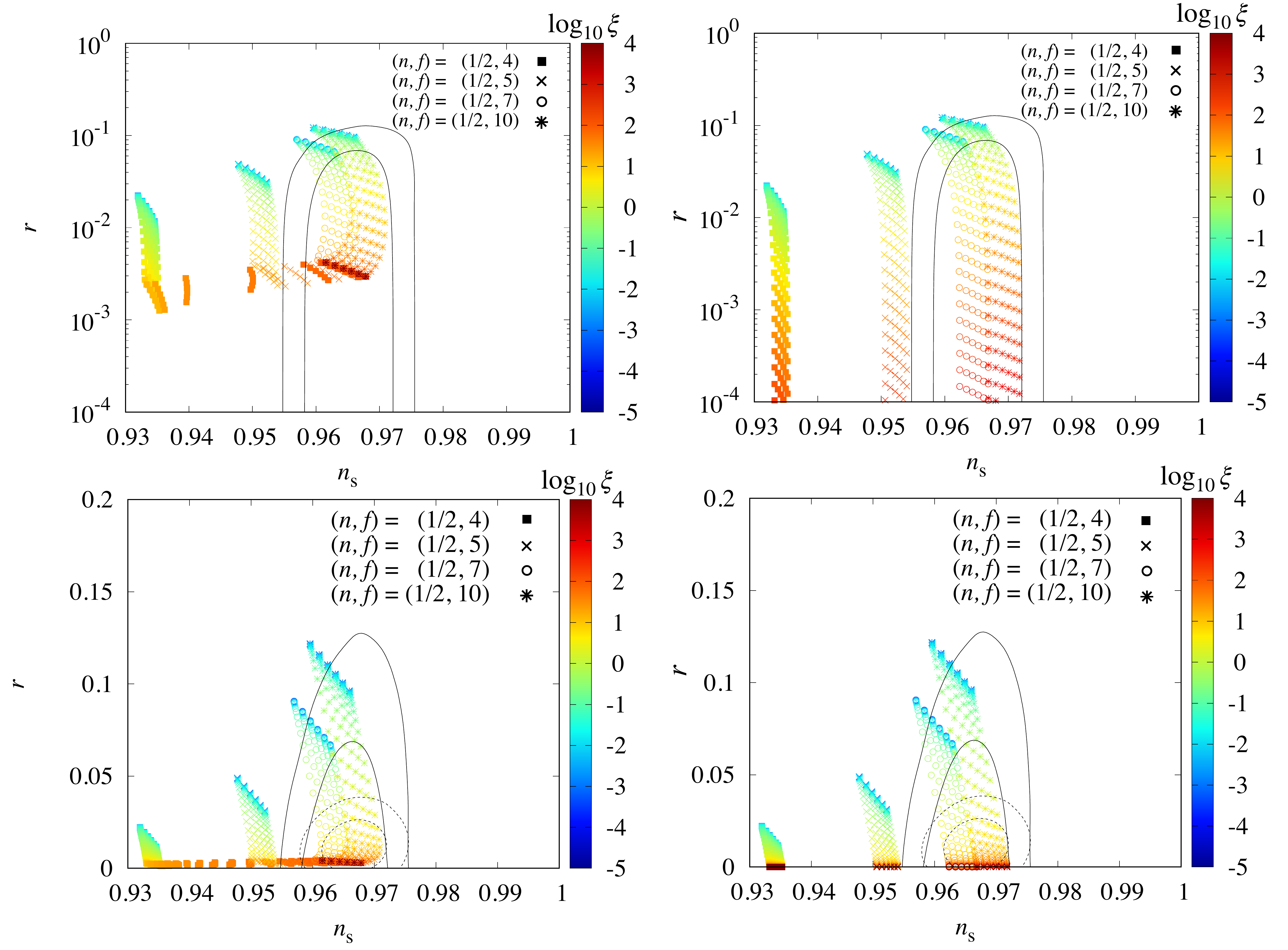}
\end{center}
\caption{\label{fig:attractor_natural}  Same as Fig.~\ref{fig:attractor_chaotic} but for the case of natural inflation with cosine-type ${\cal F}$ for $n=1/2$ which corresponds to the attractor type for several values of $f$. In the figure, the values of $f$ are shown in units of $M_{\rm pl}$.}
\end{figure}

\begin{figure}[ht]
\begin{center}
\includegraphics[width=16.5cm]{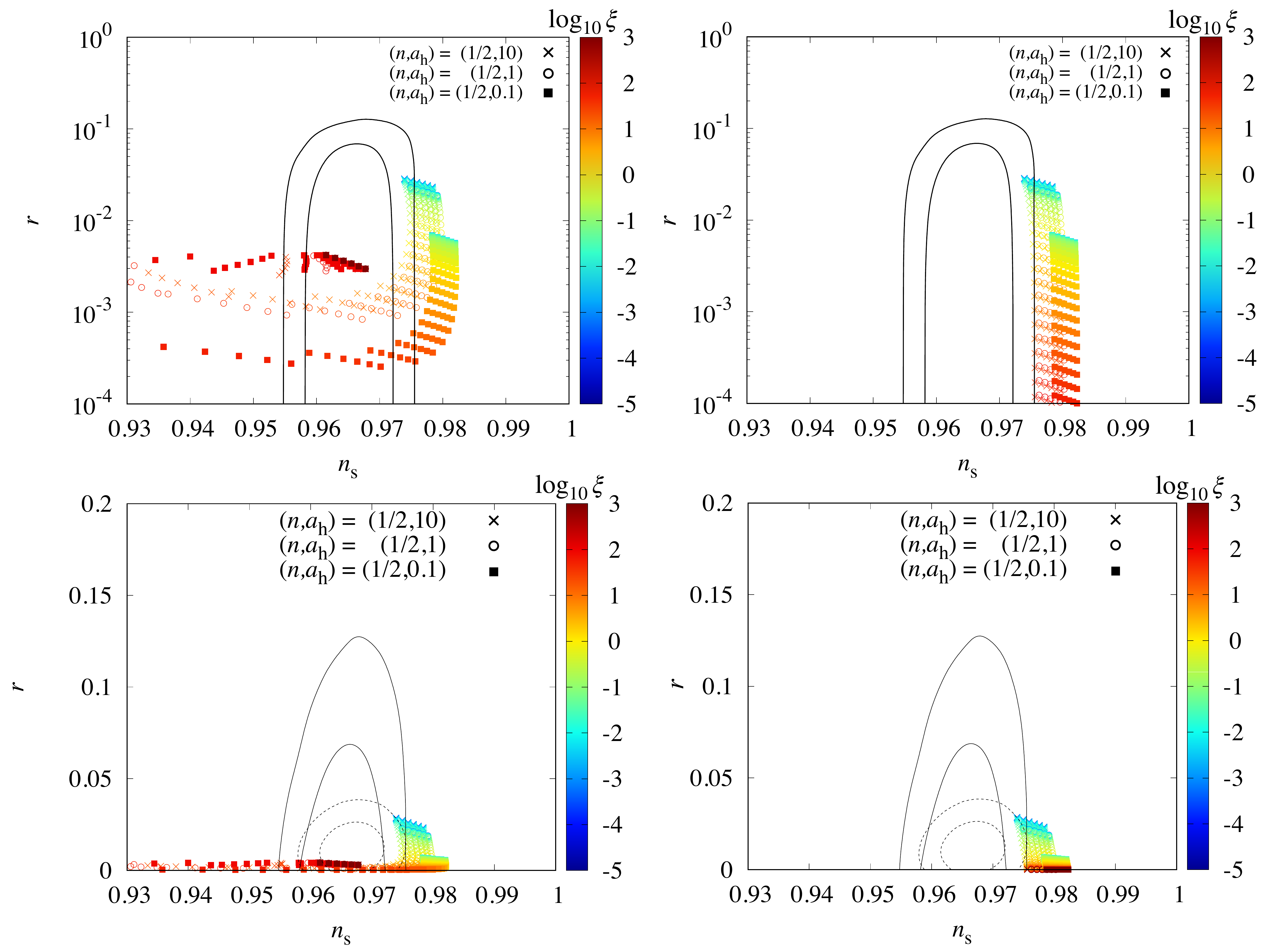}
\end{center}
\caption{\label{fig:attractor_loop}  Same as Fig.~\ref{fig:attractor_chaotic} but for the case of  loop inflation with log-type ${\cal F}$ for $n=1/2$ which corresponds to the attractor type for several values of $a_h$.}
\end{figure}

In Figs.~\ref{fig:attractor_chaotic}, \ref{fig:attractor_natural} and \ref{fig:attractor_loop},  we respectively show the cases of chaotic inflation with power-law ${\cal F}$ for some values of $p$ and $n$,  natural inflation with cosine-type ${\cal F}$ for $n=1/2$ with several values of $f$,  loop inflation with log-type of ${\cal F}$ for some values of $a_h$,  all of which correspond to the attractor type.  In each figure, we show the metric (left panels) and Palatini cases (right panels) with log-scale (top panels) and linear-scale (bottom panels) for $r$. The number of $e$-folds is assumed as $N = 50-60$ for all cases.  From the figures, one can see that all models converge to the attractor prediction given in Eqs.~\eqref{eq:attractor_ns} and \eqref{eq:attractor_r} in the $n_s$-$r$ plane for the metric formulation case although the trajectories from a small $\xi$ region to the attractor one ($\xi \gg 1$) can be non-trivial in some models, which is not traceable in the analytic approach here.   On the other hand,  in the Palatini case, $r$ just decreases,  and $n_s$ is given by Eq.~\eqref{eq:ns_attractor_palatini}, which depends on the Jordan frame potential. 

The attractor predictions for $n_s$ and $r$ are well within the current constraint from Planck+BICEP/Keck~2018 \cite{BICEP:2021xfz}, and thus  inflation models can become viable by assuming a large non-minimal coupling parameter $\xi$ and the functional form of ${\cal F}$ satisfying the relation \eqref{eq:attractor_type} in the metric case even if the minimally coupled counterpart is excluded by the data. Some models can become viable also in the Palatini case as far as the original Jordan frame potential gives the spectral index, expressed as Eq.~\eqref{eq:ns_attractor_palatini},  consistent with observational bounds. The tensor-to-scalar ratio is much suppressed in the large $\xi$ limit, and then eventually becomes consistent with the constraints.

\subsection{$V_J$-dominant type}

Next we discuss the inflationary predictions for the $V_J$-dominant type. When $\xi$ is large ($\xi \gg {\cal O}(1)$), we can approximate $A(\phi) \simeq \xi {\cal F} (\phi)$. In this case, the slow-roll parameters can be written as 
\begin{eqnarray}
\label{eq:eps_large_xi}
 \epsilon & =& 
P^2 \left[ 
\epsilon_J  - 2 M_{\rm pl}^2 \frac{{\cal F}'}{\cal F} \frac{V_J'}{V_J} + 2 M_{\rm pl}^2  \left( \frac{{\cal F}'}{\cal F} \right)^2 
\right]  \notag \\
&=& P^2 \epsilon_J (1-\alpha)^2 \,, 
\end{eqnarray}
\begin{eqnarray}
\label{eq:eta_large_xi}
\eta &=& 
P^2 
\left[ 
\eta_J 
- 4 M_{\rm pl}^2 \frac{{\cal F}'}{\cal F} \frac{V_J'}{V_J}
+ 6 M_{\rm pl}^2  \left( \frac{{\cal F}'}{\cal F} \right)^2 
- 2  M_{\rm pl}^2   \frac{{\cal F}^{''}}{\cal F} 
+   M_{\rm pl}^2 \frac{P'}{P}\left( - 2 \frac{{\cal F}'}{\cal F}  + \frac{V_J'}{V_J} \right)
\right]  \notag \\
&=& P^2 \left[ 
(1-\alpha) \eta_J - 2 \alpha (1 -\alpha) \epsilon_J 
+ \frac12 \alpha (1-\alpha) P^2 \epsilon_J \left\{ \frac{1}{\xi {\cal F}} + 3 \kappa \left( \alpha \epsilon_J - \frac{\alpha}{2} \eta_J \right) \right\}
\right] \,, \notag \\  
\end{eqnarray}
and the number of $e$-folds is given by
\begin{eqnarray}
N &=& 
\frac{1}{\Mpl^2} \int_{\phi_{\rm end}}^{\phi_\ast}  \frac{1}{P^2} \left( \frac{V_J'}{V_J} - 2  \frac{\cal F'}{\cal F}  \right)^{-1} d\phi \notag \\
&=&
\frac{\alpha}{2 (1-\alpha)\Mpl^2} \int_{\phi_{\rm end}}^{\phi_\ast} \frac{1}{P^2} \frac{\cal F}{\cal F'} \, d\phi \,,
\end{eqnarray}
where we have used the relation~\eqref{eq:VJ_F_alpha}. Here $P^2$ is given by, for $\xi \gg {\cal O}(1)$, 
\begin{equation}
\label{eq:P_large_xi}
P^2 \simeq \frac{1}{ \displaystyle\frac{1}{\xi \cal F}  + \frac{3}{2} \kappa \Mpl^2\left( \frac{\cal F'}{\cal F} \right)^2} \equiv \frac{1}{ a(\phi) + b(\phi)} \,.
\end{equation}
Here we have introduced the functions $a(\phi)$ and $b(\phi)$ which are defined as 
\begin{equation}
\label{eq:a_def}
a( \phi) \equiv \frac{1}{\xi \cal F}, 
\qquad 
b(\phi) \equiv \frac{3}{2} \kappa \Mpl^2 \left( \frac{\cal F'}{\cal F} \right)^2 \,.
\end{equation}

When we discuss the observables such as $n_s$ and $r$, we fix the number of $e$-folds $N$ corresponding to the epoch when the mode of the reference scale exited the horizon.  When $\xi$ is varied, the Einstein frame potential is modified and then $\phi_\ast$ is also changed for a fixed $N$. Therefore $\phi_\ast$ implicitly depends on $\xi$, which should be kept in mind in the following argument.

To characterize the relative size of $a(\phi)$ and $b(\phi)$, we also introduce the quantity $R$ defined as
\begin{equation}
R \equiv \frac{b(\phi_\ast)}{a(\phi_\ast)} =  \left. \frac{3}{2} \kappa \xi \Mpl^2 {\cal F}  \left( \frac{\cal F'}{\cal F} \right)^2 \right|_{\phi = \phi_\ast} \,,
\end{equation}
in which $R$ is to be evaluated at $\phi = \phi_\ast$ since the observables or the slow-roll parameters are eventually calculated at the reference scale. Depending on the size of $R$, we can categorize the $V_J$-dominant type further into two classes: Case~(i) $(R  > 1)$  and Case~(ii) $(R < 1)$.   Notice that, by definition, the Palatini case only includes Case~(ii) since it  corresponds to $\kappa =0$,  and then $b(\phi)=0$.

\begin{figure}[t]
\begin{center}
\includegraphics[width=16.cm]{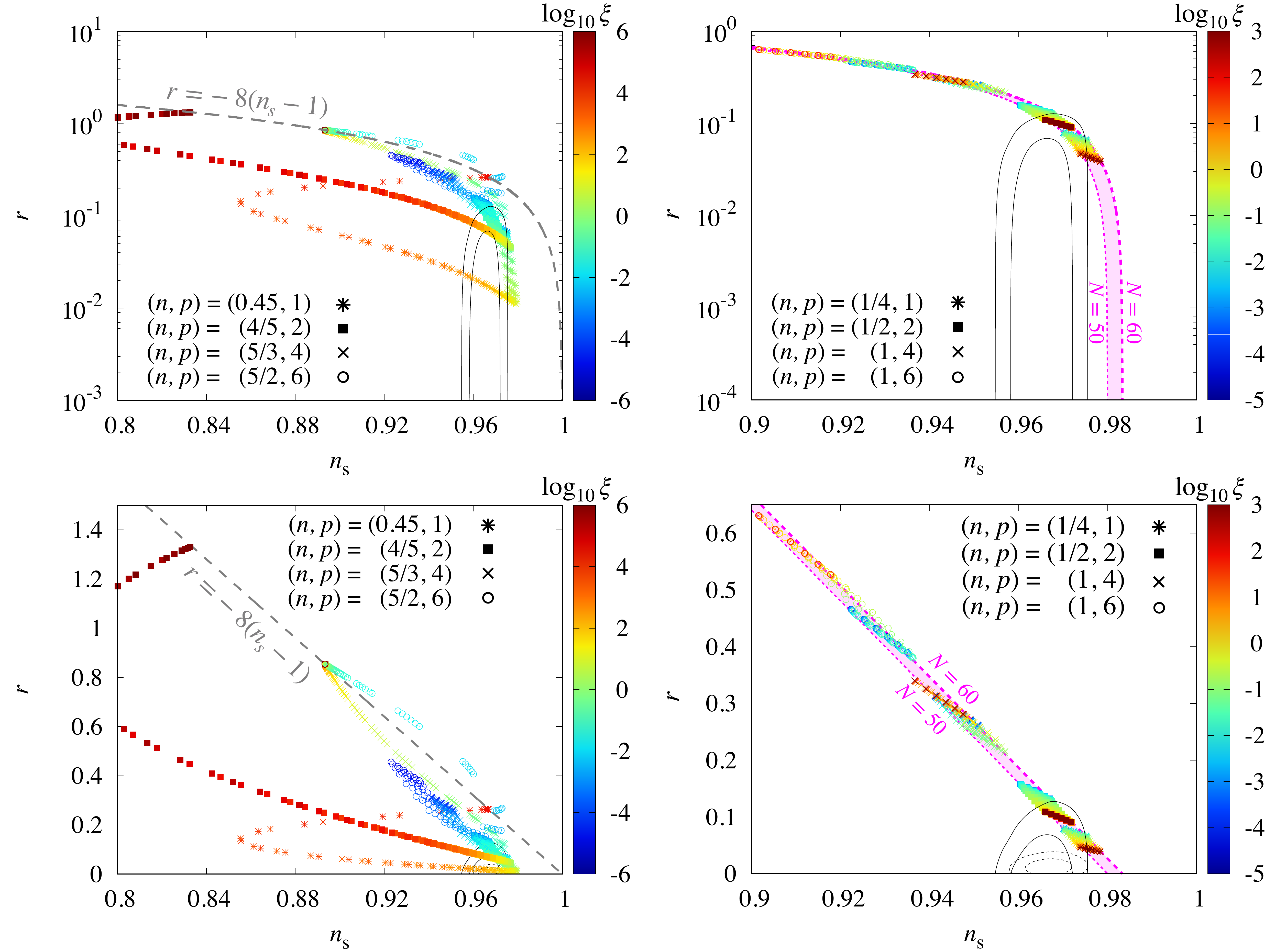}
\end{center}
\caption{\label{fig:V_dominant_chaotic}   Same as Fig.~\ref{fig:attractor_chaotic} except that the cases of chaotic inflation with power-law ${\cal F}$ are shown for Case~(i) (left column)  and Case~(ii) (right column) of the $V_J$-dominant type. In the left column, the relation~\eqref{eq:VJ_case1_consistency} for Case~(i) is also shown. In the right column, the Case~(ii) consistency relation given in Eq.~\eqref{eq:ns_r_VJdom_case2_chaotic} for $N=50$ and $60$ are depicted.}
\end{figure}

First we discuss Case~(i) in which $R >1$ holds. In this case, $P^2$ can be approximated as
\begin{equation}
P^2 = \frac{1}{ \displaystyle\frac{3}{2}\Mpl^2 \left( \frac{\cal F'}{\cal F} \right)^2} \,,
\end{equation}
where we have neglected $a(\phi)$ in the denominator in Eq.~\eqref{eq:P_large_xi} since $a(\phi) < b(\phi)$, and then $\epsilon$ and $\eta$ are reduced to the following form:
\begin{eqnarray}
&& \epsilon = \frac43 \left( \frac{\alpha -1}{\alpha} \right)^2  \,,
\qquad\qquad 
\eta = \frac83 \left( \frac{\alpha -1}{\alpha} \right)^2  \,.
\end{eqnarray}
These expressions are obtained by neglecting  the term with $1 / (\xi \cal F)$ in the last curly bracket in Eq.~\eqref{eq:eta_large_xi}.  As one can see from the above formulas, the slow-roll parameters converge to some certain values as $\xi$ gets large for a fixed $\alpha$.  One can also notice that $\epsilon$ and $\eta$ are related as 
\begin{equation}
\eta = 2 \epsilon \,.
\end{equation}
Therefore $n_s$ and $r$ are predicted, for sufficiently large $\xi$,  as
\begin{equation}
n_s =  1 - \frac83 \left( \frac{\alpha -1}{\alpha} \right)^2 \,,
\qquad
r = \frac{64}{3} \left( \frac{\alpha -1}{\alpha} \right)^2 \,.
\end{equation}
Interestingly,  these expressions lead to the consistency relation for $n_s$ and $r$:
\begin{equation}
\label{eq:VJ_case1_consistency}
r = -8 (n_s -1) \,. 
\end{equation}
Actually models satisfying this relation have already been excluded by Planck data. In Fig.~\ref{fig:V_dominant_chaotic},  this relation is depicted with gray dashed line along with a $V_J$-dominant case example for chaotic inflation with power-law ${\cal F}$. As clearly seen from the figure, the line is well above the Planck constraint and thus models of this class  are excluded at least at large $\xi$. The predictions for chaotic inflation with power-law ${\cal F}$ for the $V_J$-dominant case will be discussed later in this section.

Next we consider Case (ii) in which $R <1$ holds. In this case, $P^2$ can be approximated as 
\begin{equation}
P^2 = \xi \cal F \,,
\end{equation}
where we have neglected the second term in the denominator of Eq.~\eqref{eq:P_large_xi}. Approximately we can also keep only the first term  in the last curly bracket in Eq.~\eqref{eq:eta_large_xi} for $\eta$, then we obtain the slow-roll parameters in this case as 
\begin{eqnarray}
\epsilon &=&  (1-\alpha )^2 \epsilon_J {\cal F} \xi \,, \\ [8pt]
\qquad\qquad 
\eta &=& {\cal F} \xi \left[ 
(1-\alpha) \eta_J - \frac32 \alpha (1-\alpha) \epsilon_J 
\right] \,.
\end{eqnarray}
Therefore $n_s$ and $r$ in Case (ii) are given by 
\begin{eqnarray}
\label{eq:ns_VJdom_2}
n_s -1 & = &  3 {\cal F}\xi (1-\alpha) (2 -\alpha) \epsilon_J \left[ X -1 \right] \,,  \\ [8pt] 
r & = & 16 (1-\alpha )^2 \epsilon_J {\cal F} \xi \,,
\end{eqnarray}
where we have defined 
\begin{equation}
\label{eq:def_X}
X \equiv \frac{2}{3(2-\alpha)} \frac{\eta_J}{\epsilon_J} \,.
\end{equation}
From the expression~\eqref{eq:ns_VJdom_2}, when $X > 1$, or  the inequality
\begin{equation}
\frac{\eta_J}{\epsilon_J} > \frac{2}{3(2-\alpha)} 
\end{equation}
is satisfied, the spectral index is larger than unity, and hence such models are excluded by current observations.

\begin{figure}[t]
\begin{center}
\includegraphics[width=16.5cm]{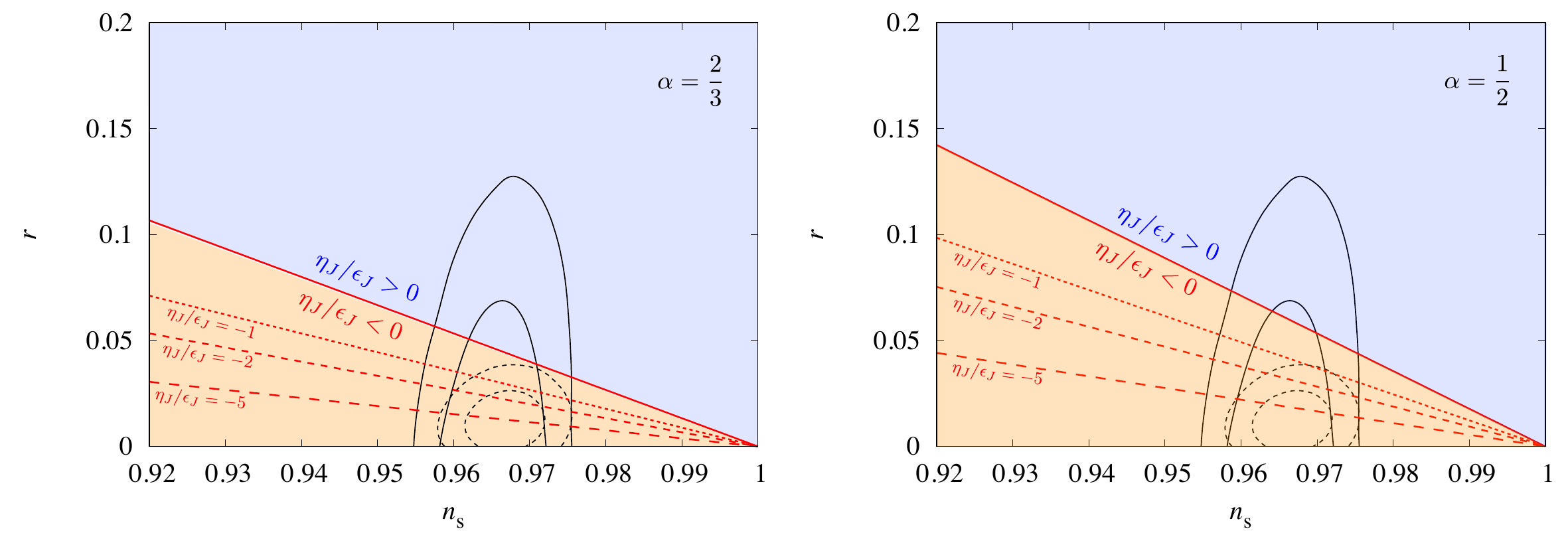}
\end{center}
\caption{\label{fig:V_dominant_case2}  Predictions of $n_s$ and $r$ in Case~(ii) of the $V_J$-dominant type for $\alpha = 2/3$ (left) and $\alpha=1/2$ (right).  1$\sigma$ and 2$\sigma$ constraints from Planck \cite{Planck:2018jri}  and Planck+BAO+BICEP/Keck~2018 \cite{BICEP:2021xfz} are also depicted with solid and dashed lines, respectively.}
\end{figure}

Furthermore, the expressions for $n_s$ and $r$ give the following relation:
\begin{equation}
\label{eq:VJdom_case2_ns_r_rel} 
r = - 8 (n_s -1 )\frac{{\cal A}(\alpha)}{ 1 - X} \,,
\end{equation}
where
\begin{equation}
{\cal A}(\alpha) = \frac{2(1-\alpha)}{3(2-\alpha)}  \,.
\end{equation}
Notice that $\xi$ does not appear in Eq.~\eqref{eq:VJdom_case2_ns_r_rel}  although the values of $\epsilon_J$ and $\eta_J$ implicitly depend on $\xi$ since the value of $\phi_\ast$ changes as $\xi$ varies when we fix the number of $e$-folds as $N =50-60$. From the relation~\eqref{eq:VJdom_case2_ns_r_rel}, one can see that, depending on $\alpha$ and the ratio of the Jordan frame slow-roll parameters $\eta_J / \epsilon_J$, some regions in the  $n_s$--$r$ plane cannot be reached for $\xi \gg {\cal O}(1)$.

In Fig.~\ref{fig:V_dominant_case2}, the cases of $\alpha = 2/3$ and $1/2$ are shown. When $\alpha = 2/3$,  if the Jordan frame slow-roll parameters $\eta_J $ is positive\footnote{
Although the actual condition here is $X > 0$,   since $\epsilon_J$ is positive by definition and $ 0 < {\cal A}(\alpha) < 1/3$ holds in the $V_J$-dominant case, the condition $\eta_J >0 $ is sufficient in the subsequent argument.
}, the values of $n_s$ and $r$ are always above the line of $r = -(4/3) (n_s - 1)$, which is represented by blue region  in the left panel of Fig.~\ref{fig:V_dominant_case2}.  The figure indicates that this region is incompatible with the current constraint from Planck+BAO+BICEP/Keck~2018  \cite{BICEP:2021xfz}, which means that when $\alpha = 2/3$ and  $\eta_J >0$,  the model is excluded for $\xi \gg {\cal O}(1)$ for any models with $V_J$ and ${\cal F}$ as far as the model is classified as Case~(ii) in the $V_J$-dominant type.  We should emphasize that this argument only depends on the value of $\alpha$ and  the sign of $\eta_J$. For the case of $\alpha = 1/2$, when $\eta_J / \epsilon_J > -1$, such models are not allowed by the current constraint, which can be seen from the right panel of Fig.~\ref{fig:V_dominant_case2}.  

It should be noted that even if $\eta_J < 0$ for $\alpha= 2/3$, which is shown by the orange  region in Fig.~\ref{fig:V_dominant_case2}, it does not necessarily indicate that models can become viable for $\xi \gg {\cal O}(1)$. Since the values of $n_s$ and $r$ should be compared to observational constraints at the reference scales,  one needs to relate the value of $N$ and $\phi_\ast$,  with which $\epsilon_J, \eta_J$ and ${\cal F}$ are evaluated. The number of $e$-folds in Case~(ii) is given by 
\begin{equation}
\label{eq:N_VJ_case2}
N = \frac{\alpha}{2(1-\alpha) \xi\Mpl^2} \int_{\phi_{\rm end}}^{\phi_\ast} \frac{d\phi}{\cal F'} \,,
\end{equation}
from which one can relate $\phi_\ast$ and $N$. 

In the following, we consider explicit examples listed in Sec.~\ref{sec:example_models} to discuss the predictions in the $V_J$-dominant type, based on the general argument above. 

\subsubsection*{Example: Chaotic inflation with power-law ${\cal F}$}
In this model, $R$ is given by 
\begin{equation}
R = \frac{3 n^2 \xi}{2} \left( \frac{\phi_\ast}{\Mpl} \right)^{n-2} \,.
\end{equation}
Since in chaotic inflation model, $\phi_\ast$ tends to be larger than $\Mpl$ for $N=50-60$ even with a non-minimal coupling, and hence, when $n > 2$, $R$ is generally larger than 1  for $\xi \gg 1$, which is classified as Case~(i). On other hand, Case~(ii) arises when $n<2$,  although even with $n<2$, $R$ can be larger than unity for large $\xi$ and classified as Case~(i), whose examples are given in the left panel of Fig.~\ref{fig:V_dominant_chaotic}.

In Fig.~\ref{fig:V_dominant_chaotic}, the predictions in the $n_s$--$r$ plane for the $V_J$-dominant type with the chaotic inflation are shown for several sets of $p$ and $n$ whose values are given in the figure.  In the left and right panels, models corresponding to Case~(i) and (ii) are respectively depicted. As shown in the left panel, the predictions for $n_s$ and $r$ approach  some points on the line corresponding to Eq.~\eqref{eq:VJ_case1_consistency}. Although, as already mentioned, the line represented by Eq.~\eqref{eq:VJ_case1_consistency} is well above the current constraint, the trajectories in the $n_s$--$r$ plane followed by varying $\xi$ are somewhat non-trivial, and hence the model may be allowed by observational constraint at some intermediate values of $\xi$.  However, this is only possible in some limited range of $\xi$, and the general trend is that models of Case~(i) would not be viable with respect to current observational constraints.

For Case~(ii), the number of $e$-folds for $n \ne 2$  can be calculated as 
\begin{equation}
\label{eq:Ne_VJ_case2_chaotic}
N \simeq \frac{1}{( p-2n) (2-n) \xi }  \left( \frac{\phi_\ast}{\Mpl} \right)^{2-n} \,.
\end{equation}
By using this expression, one can rewrite $\phi_\ast$ as a function of $N$, from which  $n_s$ and $r$ are given by 
\begin{eqnarray}
n_s -1 &=& \frac{1}{(2-n)N} \left[-p + 3n -2 \right] \,, \\  [8pt]
r &=& \frac{ 8(p-2n)}{2-n} \frac{1}{N} \,. 
\end{eqnarray}
Actually, from these expressions, one can derive the relation between $n_s$ and $r$ as 
\begin{eqnarray}
\label{eq:ns_r_VJdom_case2_chaotic}
r &=& - 8 (n_s -1) - \frac{8}{N} \,.
\end{eqnarray}
This relation holds for any value of $n$ and $p$, except for the case with $n=2$ since Eq.~\eqref{eq:Ne_VJ_case2_chaotic} is derived under the assumption of $n\ne 2$. In the right panel of Fig.~\ref{fig:V_dominant_chaotic}, several examples for Case~(ii) are shown along with the line for the relation~\eqref{eq:ns_r_VJdom_case2_chaotic}. As seen from the figure, the predictions in Case~(ii) approach to the ones given by Eq.~\eqref{eq:ns_r_VJdom_case2_chaotic} as $\xi$ increases. For $N=50-60$, the predicted values from Eq.~\eqref{eq:ns_r_VJdom_case2_chaotic} are outside the current constraint, and thus the chaotic inflation with power-law ${\cal F}$ cannot become viable for Case~(ii) of the $V_J$-dominant type.

\subsubsection*{Example: Natural inflation with cosine-type ${\cal F}$}
In this model $R$ parameter is given by
\begin{equation}
R = \frac32 \kappa \Mpl^2 \xi \left( \frac{\Mpl}{f} \right)^2 n^2 \left( 1 - \cos \left( \frac{\phi_\ast}{f} \right) \right)^{n-2} \sin^2 \left( \frac{\phi_\ast}{f} \right) \,.
\end{equation}
As $\xi$ increases, the value of $\phi_\ast$ approaches to $\pi f$ to obtain the amount of the $e$-folds as $N = 50 -60$. By expanding $R$ at around $\phi_\ast / f = \pi$, one obtains 
\begin{equation}
R \simeq \frac32 \kappa \Mpl^2 \xi \left( \frac{\Mpl}{f} \right)^2 n^2 \, 2^{n-2} \left( \pi -  \frac{\phi_\ast}{f} \right)^2 \,,
\end{equation}
from which one can see that,  even when $\xi$ is large, the suppression due to $\left( \pi -  \phi_\ast / f \right)^2$ drives $R$ to be less than unity. Therefore the natural inflation with cosine-type ${\cal F}$ in the $V_J$-dominant case is mostly classified as Case~(ii).

\begin{figure}[ht]
\begin{center}
\includegraphics[width=16.5cm]{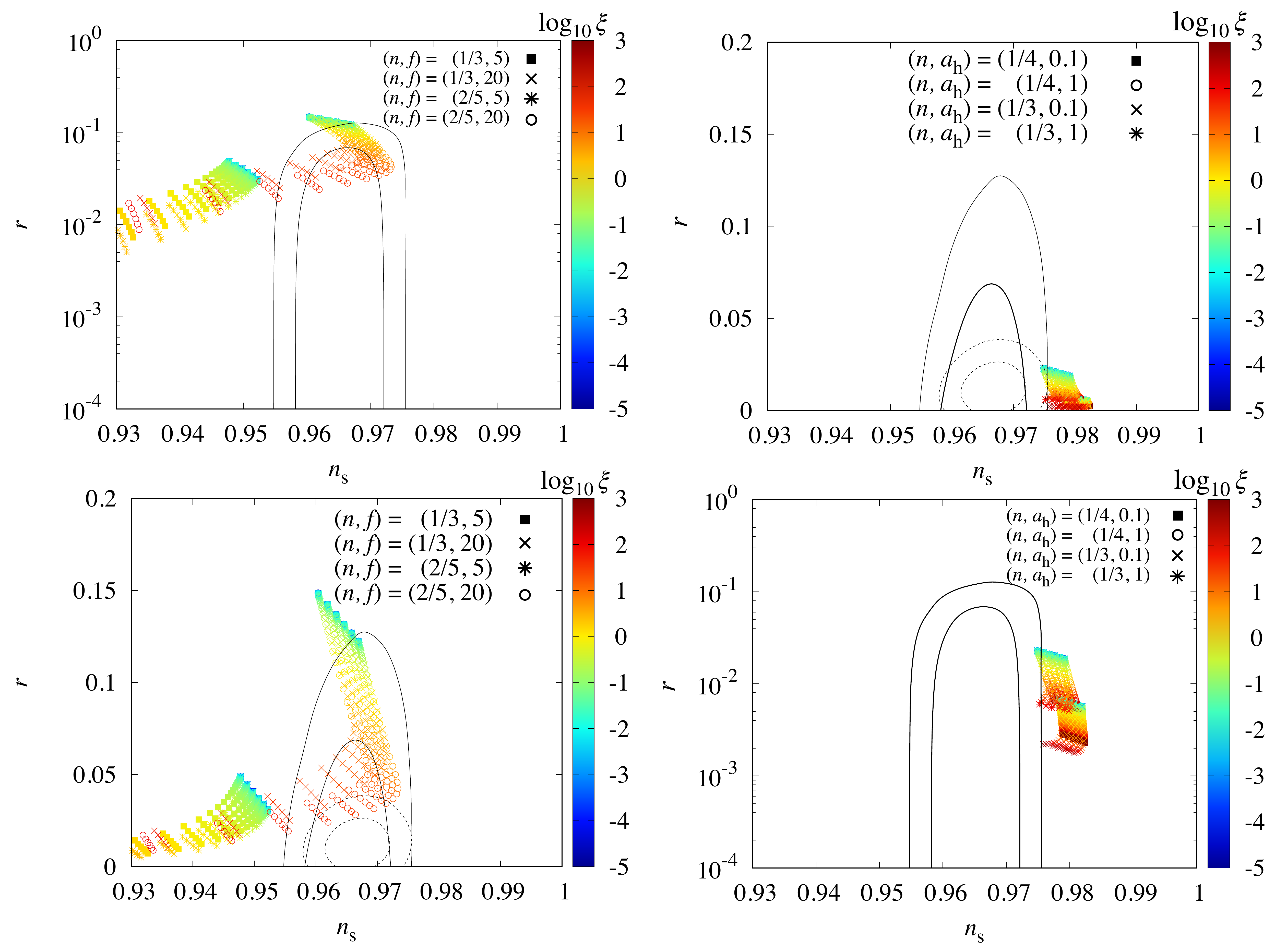}
\end{center}
\caption{\label{fig:V-dominant_natural_loop} Same as Fig.~\ref{fig:attractor_chaotic} except that the cases of natural inflation with cosine-type ${\cal F}$ (left column) and loop inflation with log-type ${\cal F}$ (right column) for Case~(ii) of  the $V_J$-dominant type are shown. }
\end{figure}

In this case, one can write down the spectral index $n_s$ and the tensor-to-scalar ratio $r$ as 
\begin{eqnarray}
n_s -1  & = &  -\xi {\cal F} \left[ 2 (6n^2 - 7n +2) \epsilon_J + (1-2n)  \left( \frac{\Mpl}{f} \right)^2  \right] \,, \\ [8pt]
r  & = &  16 (1 -2n)^2 \xi {\cal F} \epsilon_J \,.
\end{eqnarray}
As mentioned above, as $\xi$ increases, $\phi_\ast$ approaches to the value at the extremum, $\phi_\ast \sim f \pi$, and hence one can write down the Jordan frame slow-roll parameter $\epsilon_J$ by expanding around $\phi_\ast = f \pi$, at the leading order in  $  \phi_\ast / f -\pi $, as 
\begin{equation}
\epsilon_J = \frac18 \left( \frac{M_{\rm pl}}{f} \right)^2 (\phi_\ast / f -\pi)^2  \,.
\end{equation}
The value of $\eta_J$ can be obtained by utilizing the relation~\eqref{eq:eta_eps_natural}.  Then, we can express $n_s$ and $r$  as 
\begin{eqnarray}
\label{eq:ns_natural_VJ}
n_s -1  & = &  -  2^n \xi (1 - 2n)\left(\frac{\Mpl}{f}\right)^2  \,, \\
\label{eq:r_natural_VJ}
r  & = &  2^{n+1} (1-2n)^2 \xi \left( \frac{M_{\rm pl}}{f} \right)^2 (\phi_\ast / f -\pi)^2 \,. 
\end{eqnarray} 
These expressions indicate that the spectral index gets more red-tilted and the tensor-to-scalar ratio is more suppressed as $\xi$ increases.  In the left column of Fig.~\ref{fig:V-dominant_natural_loop}, the predictions for $n_s$ and $r$ in the natural inflation with cosine-type ${\cal F}$ are shown.  The behavior in the $n_s$--$r$ plane with varying $\xi$ matches well with the one indicated by the analytic expressions \eqref{eq:ns_natural_VJ} and \eqref{eq:r_natural_VJ}. In some cases, the natural inflation can become viable in the $V_J$-dominant type especially when the minimally coupled counterpart predicts larger $n_s$ and $r$ than the observational bounds. One of such examples is the case of $f=20 M_{\rm pl}$, which is depicted in Fig.~\ref{fig:V-dominant_natural_loop}.

\subsubsection*{Example: Loop inflation with log-type ${\cal F}$}
In this case, the parameter $R$ is given by 
\begin{equation}
R = \frac32 \kappa \xi n^2 a_h^2 \left( \frac{M_{\rm pl}}{\phi_\ast} \right)^2 \left[ 1+ a_h \log \left( \frac{\phi_\ast}{M_{\rm pl}} \right) \right]^{n-2} \,.
\end{equation}
Actually, this model is also mostly classified as Case~(ii) as we argue below. For Case~(ii), the number of $e$-folds is given by Eq.~\eqref{eq:N_VJ_case2}, and $N$ can be written in this model as 
\begin{equation}
\label{eq:VJ_N_loop}
N \simeq \frac{1}{2 (1-2n) \xi a_h} \left[ 1+ a_h \log \left( \frac{\phi_\ast}{M_{\rm pl}} \right) \right]^{1-n} \left( \frac{\phi_\ast}{M_{\rm pl}} \right)^2  \,,
\end{equation}
where we have approximated $ \log (\phi / M_{\rm pl})$ as a constant when we performed the integral and assumed $\phi \simeq \phi_\ast$ since this term is slowly varying. Then $R$ can be rewritten as 
\begin{equation}
R \simeq  \frac{3 \kappa n^2}{4 (1-2n) N} \left( \frac{a_h}{ 1+ a_h \log (\phi_\ast / M_{\rm pl} )}\right) \,.
\end{equation}
Since $0< 2n  <1$ in the $V_J$-dominant type, the above expression indicates that $R$ would be less than unity since $\phi_\ast > M_{\rm pl}$ when $\xi$ is large as can be implied by Eq.~\eqref{eq:VJ_N_loop}.  Thus this model can also be mostly classified as Case~(ii) for the $V_J$-dominant type.

By using Eq.~\eqref{eq:VJ_N_loop}, one can express $\epsilon_J$ with $N$ as 
\begin{equation}
\epsilon_J \simeq \frac{a_h}{ 4 (1-2n) \xi  N}  \left[ 1+ a_h \log \left( \frac{\phi_\ast}{M_{\rm pl}} \right) \right]^{-n -1} \,.
\end{equation}
Then the spectral index and the tensor-to-scalar ratio can be written as 
\begin{eqnarray}
n_s -1 & \simeq & -\frac{1}{N} \left( \frac{a_h}{1+ a_h \log (\phi_\ast / M_{\rm pl})} \right)^2 
\left[1 + \frac{3 (1-n)}{2}  \frac{1+ a_h \log (\phi_\ast / M_{\rm pl} )}{a_h} \right]  \,, \\ [8pt]
r & \simeq &  \frac{4(1-2n)}{N} \frac{a_h}{1+ a_h \log (\phi_\ast / M_{\rm pl} )} \,.
\end{eqnarray}
Since the term with $ \log (\phi_\ast / M_{\rm pl} )$ increases as $\xi$ gets larger (although the change is mild), $r$ is suppressed. The spectral index is also slightly changed, but it reacts weakly  against the change of $\xi$ since $n_s$ depends on $\xi$ only indirectly through $\log ( \phi_\ast /M_{\rm pl})$. In the right column of Fig.~\ref{fig:V-dominant_natural_loop}, several examples for the loop inflation with log-type ${\cal F}$ are shown.  As expected, the changes of $n_s$ and $r$ for varying $\xi$ are mild.  Since the minimally coupled version of this model already gives a small $r$, the size of $r$ can be consistent with current observations by  suitably choosing the value of $a_h$. However, the spectral index is somewhat larger than the current bound for the minimally coupled case. Even if we assume a large non-minimal coupling, the change of $n_s$ is not so significant in the $V_J$-dominant type, and thus this model would not become viable when we choose ${\cal F}$ categorized as the $V_J$-dominant type.

\subsection{${\cal F}$-dominant type}

As  discussed in Sec.~\ref{sec:classification}, the Einstein frame potential in the ${\cal F}$-dominant type has an extremum even if the minimally coupled  counterpart does not have it.  When the value of $\chi$ ($\phi$)  exceed the extremum value $\chi_{\rm ex}$ ($\phi_{\rm ex}$) at which $ d V_E /d \chi =0$ is satisfied, the inflaton runs away  to a large value of $\chi$ ($\phi$), and hence we do not consider such a case. 

When we limit ourselves to the range of $\chi$ ($\phi$) as $\chi < \chi_{\rm ex}$ ($\phi < \phi_{\rm ex}$), $\chi_\ast$ ($\phi_\ast$) needs to be close to $\chi_{\rm ex}$ ($\phi_{\rm ex}$) to obtain a sufficient amount of $e$-folds as $N=50-60$.  Therefore,  in this case, the slow-roll parameter $\epsilon$ at $\chi_\ast$ can be expanded around  $\chi_{\rm ex}$ as 
\begin{equation}
\epsilon (\chi_\ast) = \epsilon (\chi_{\rm ex}) + \left. \frac{d\epsilon}{d\chi} \right|_{\chi = \chi_{\rm ex}}  (\chi_\ast - \chi_{\rm ex}) 
+ \frac12 \left. \frac{d^2\epsilon}{d\chi^2} \right|_{\chi = \chi_{\rm ex}}  (\chi_\ast - \chi_{\rm ex})^2 + {\cal O} \left(  (\chi_\ast - \chi_{\rm ex})^3 \right) \,.
\end{equation}
Since $\chi_{\rm ex}$ satisfies  $ d V_E /d \chi =0$, the first order term in the above expression vanishes. Furthermore, the second derivative of $\epsilon$ with respect to $\chi$ is given by 
\begin{eqnarray}
\left.  \frac{d^2\epsilon}{d\chi^2} \right|_{\chi = \chi_{\rm ex}}  &=&
\left. \frac12  \left(
 \frac{2}{V_E^2} \left( \frac{d^2 V_E}{d\chi^2} \right)^2 -  \frac{1}{V_E^2} \frac{d V_E}{d\chi} \frac{d^3 V_E}{d\chi^3}  
 -\frac{10}{V_E^3} \left( \frac{d V_E}{d\chi} \right)^2 \frac{d^2 V_E}{d\chi^2}
 -\frac{6}{V_E^4}   \left( \frac{d V_E}{d\chi} \right)^4
 \right)  \right|_{\chi = \chi_{\rm ex}} 
   \notag \\ [8pt]
 &=& 
\left.  \frac{1}{V_E^2} \left( \frac{d^2 V_E}{d\chi^2} \right)^2  \right|_{\chi = \chi_{\rm ex}}
= \eta^2 (\chi_{\rm ex}) \,,
\end{eqnarray}
where we have used the fact that $\epsilon (\chi_{\rm ex})=0$ since $ d V_E /d \chi |_{\chi = \chi_{\rm ex}} =0$.  Therefore $\epsilon (\chi_\ast)$ is written, at the leading order in $\chi_\ast - \chi_{\rm ex}$,  as
\begin{equation}
\epsilon (\chi_\ast) =  \frac12 \eta^2 (\chi_{\rm ex})  (\chi_\ast - \chi_{\rm ex})^2  \,, 
\end{equation}
which indicates that $\epsilon$ is given by the second order in $\eta$. In this case,  the tensor-to-scalar ratio becomes
\begin{equation}
\label{eq:Fdom_r}
r \simeq 8 \eta^2 (\chi_{\rm ex})  (\chi_\ast - \chi_{\rm ex})^2 \,.
\end{equation}
From this expression, one can see that $r$ would be very suppressed since it is proportional to $\eta^2$ and $(\chi_\ast - \chi_{\rm ex})^2$,  both of which are the square of a small quantity.  Since $\epsilon$ is given by the second order in $\eta$, we can write the spectral index as
\begin{equation}
\label{eq:Fdom_ns}
n_s - 1 \simeq 2  \eta (\chi_{\rm ex}) \,.
\end{equation}
The extremum we consider here  corresponds to the local maximum where $d^2V_E / d\chi^2 < 0$ holds,  and thus $\eta (\chi_{\rm ex})$ is negative, which means that the spectral index becomes red-tilted. To represent $\eta(\chi_{\rm ex})$ by $\phi_{\rm ex}$, first we rewrite Eq.~\eqref{eq:eta_general} using Eq.~\eqref{eq:VJ_F_alpha},
\begin{align}
\label{eq:eta_Fdom}
\eta(\phi) = P^2(\phi)\left[
\al\left\{
\frac{3\al}{(1 + a(\phi))^2} - \frac{\al + 2}{1 + a(\phi)}
\right\}\ep_J(\phi) + \left(
\eta_J(\phi) + \Mpl^2 \frac{P^\pr}{P} \frac{V_J^\pr}{V_J}
\right) \left(
1 - \frac{\al}{1 + a(\phi)}
\right)
\right]\,,
\end{align}
where $a(\phi)$ is the function defined in Eq.~\eqref{eq:a_def}.  At the extremum, 
\begin{equation}
\label{eq:extremum}
\frac{d V_E }{d\chi }  = \frac{d\phi}{d\chi} \frac{V_J}{A^2}  \left( \frac{V_J'}{V_J}  - 2  \frac{A'}{A} \right) =0 
\end{equation}
 holds, which gives  
\begin{equation}
\label{eq:a_alpha_Fdom}
a(\phi_{\rm ex}) = \al - 1 \,,
\end{equation}
By using this relation,  $\eta_J(\phi_{\rm ex})$ can be rewritten by a very simple form:
\begin{align}
\label{eq:eta_Fdom_ex}
\eta(\phi_{\rm ex}) = - \left(
\frac{\al- 1}{\al} + \frac{3}{4} \kappa \ep_J(\phi_{\rm ex})
\right)^{-1} 
 (\al- 1 )\ep_J(\phi_{\rm ex})\,.
\end{align}
Since $\al > 1$ in the ${\cal F}$-dominant type and $\epsilon_J >0$ by definition, the above expression indicates that  $\eta(\phi_{\rm ex}) < 0$ always holds.  

The value of $\chi_{\rm ex}$ gets smaller as $\xi$ increases, which can be seen from the figures of the Einstein frame potential shown in Figs.~\ref{fig:VE_chaotic}, \ref{fig:VE_natural} and \ref{fig:VE_loop}. Since we take that $d \chi / d\phi >0$, $\phi_{\rm ex}$ also becomes smaller as $\xi$ gets larger.  In addition, we assume that the Jordan frame potential satisfies $d V_J / d\phi >0$ for the range of $\phi$ relevant to the inflationary dynamics. This means that, as the value of $\phi$ becomes smaller, or as $\xi$ increases, 
$\epsilon_J$ becomes larger, which indicates that $ | \eta (\phi_{\rm ex}) |$ gets larger as $\xi$ increases. 

From the argument above,  one can see that the spectral index in the ${\cal F}$-dominant type, given as Eq.~\eqref{eq:Fdom_ns}, becomes more red-tilted as $\xi$ increases or, as $\phi_\ast$ approaches to $\phi_{\rm ex}$ ($\chi_\ast$ approaches to $\chi_{\rm ex}$). This tendency is the same both for the metric and Palatini cases. The tensor-to-scalar ratio, given in Eq.~\eqref{eq:Fdom_r}, becomes small as $\xi$ increases since $\chi_\ast$ further approaches to $\chi_{\rm ex}$. (Although $\epsilon_J$ gets larger as $\xi$ increases, the factor of $(\chi_\ast - \chi_{\rm ex})^2$ is more important in Eq.~\eqref{eq:Fdom_r}.)  In the following, we discuss how these trends in the $\cal F$-dominant type actually appear in explicit models presented in Sec.~\ref{sec:example_models}.

\begin{figure}[ht]
\begin{center}
\includegraphics[width=12cm]{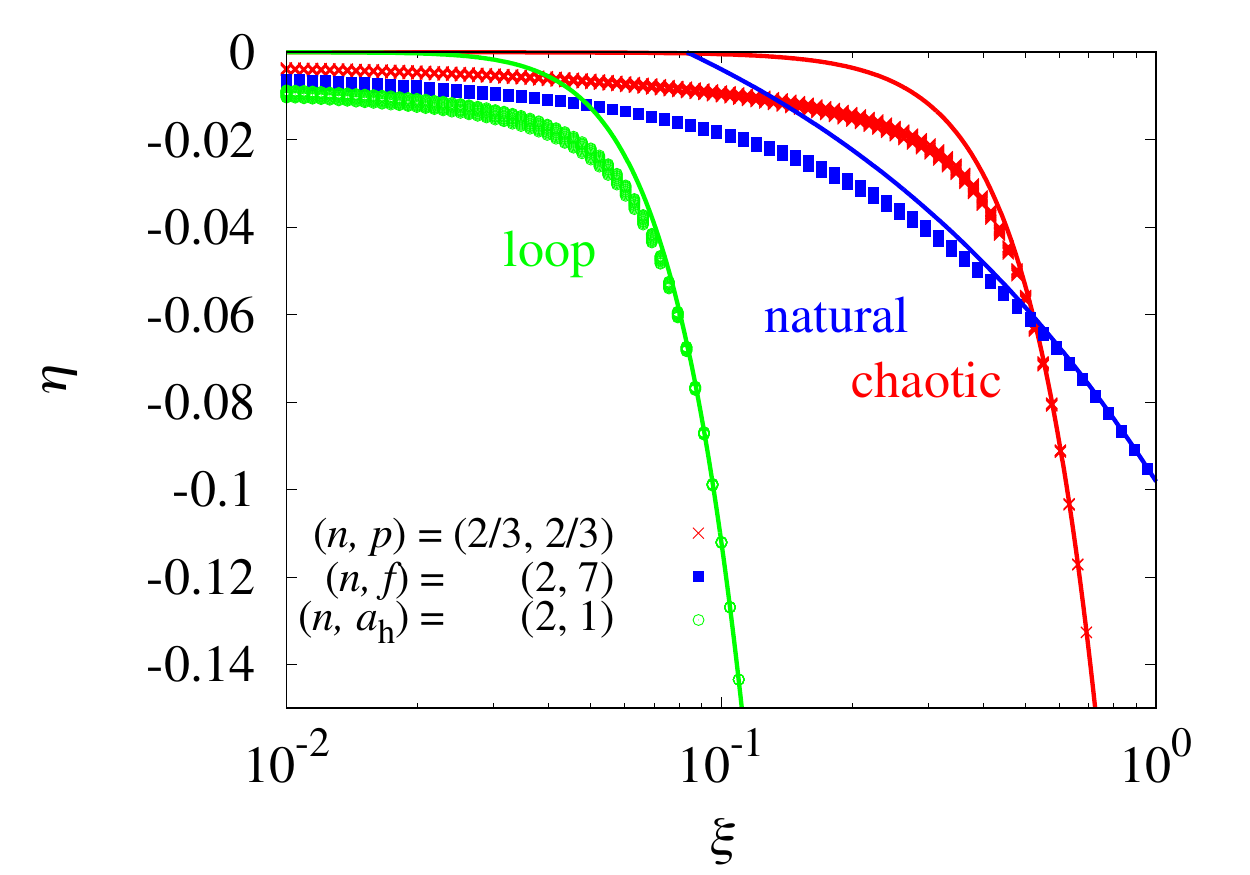}
\end{center}
\caption{\label{fig:Fdom_eta_xi}  Plot of $\eta$ as a function of $\xi$ for the cases with chaotic inflation with power-law type $\cal F$, natural inflation with cosine-type $\cal F$ and loop inflation with log-type ${\cal F}$. Dots represent the results from numerical calculations and  lines correspond to the analytic formulas given in Eqs.~\eqref{eq:xi_eta_chaotic}, \eqref{eq:xi_eta_natural} and \eqref{eq:xi_eta_loop} for each model. }
\end{figure}

\subsubsection*{Example: Chaotic inflation with power-law $\cal{F}$}

The slow-roll parameter $\epsilon_J$ in this models is given in \eqref{eq:VJ_chaotic}. At the extremum of the Einstein potential, $a(\phi_{\rm ex})$ satisfies the relation~\eqref{eq:a_alpha_Fdom}, from which we obtain
\begin{align}
a(\phi_{\rm ex}) 
= \al - 1 = \frac{2n - p}{p}\,.
\end{align} 
From the condition of the extremum~\eqref{eq:extremum}, $\phi_{\rm ex}$  can be written as
\begin{align}
\label{eq:phi_ex_chaotic}
\frac{\phi_{\rm ex}}{\Mpl} =\left(
\frac{p}{(2n - p)\xi}
\right)^{1/n}\,.
\end{align}
By substituting this expression into $\ep_J$,  $\eta (\phi_{\rm ex})$ can be expressed as 
\begin{align}
\label{eq:xi_eta_chaotic}
\eta(\phi_{\rm ex}) = -np\left(
\frac{2n-p}{p}\xi
\right)^{2/n} \left[
1 + \frac{3\kappa}{4}\frac{np^2}{2n-p}\left(
\frac{2n-p}{p}\xi
\right)^{2/n}
\right]^{-1} \,.
\end{align}
For example,  for the metric case with $(n,p)= (2/3,2/3)$, we obtain
\begin{equation}
\label{eq:eta_analytic_chaotic}
\eta(\phi_{\rm ex}) = -\frac{4}{9} \frac{\xi^3}{1 + \displaystyle\frac{\xi^3}{3}}\,.
\end{equation}

In Fig.~\ref{fig:Fdom_eta_xi}, the value of $\eta$ is plotted as a function of $\xi$ for the metric case with $(n,p)= (2/3,2/3)$ along with the cases of other models. In the figure, $\eta$ from a numerical calculation is shown with dots and the analytic formula for large $\xi$ given in Eq.~\eqref{eq:eta_analytic_chaotic} is depicted with a line.  As expected, as $\xi$ becomes relatively large, the analytic formula matches with the numerically calculated value. One can also see that the value of $\eta$ becomes negatively larger as $\xi$ increases, which confirms the argument given above. 

\begin{figure}[ht]
\begin{center}
\includegraphics[width=16.5cm]{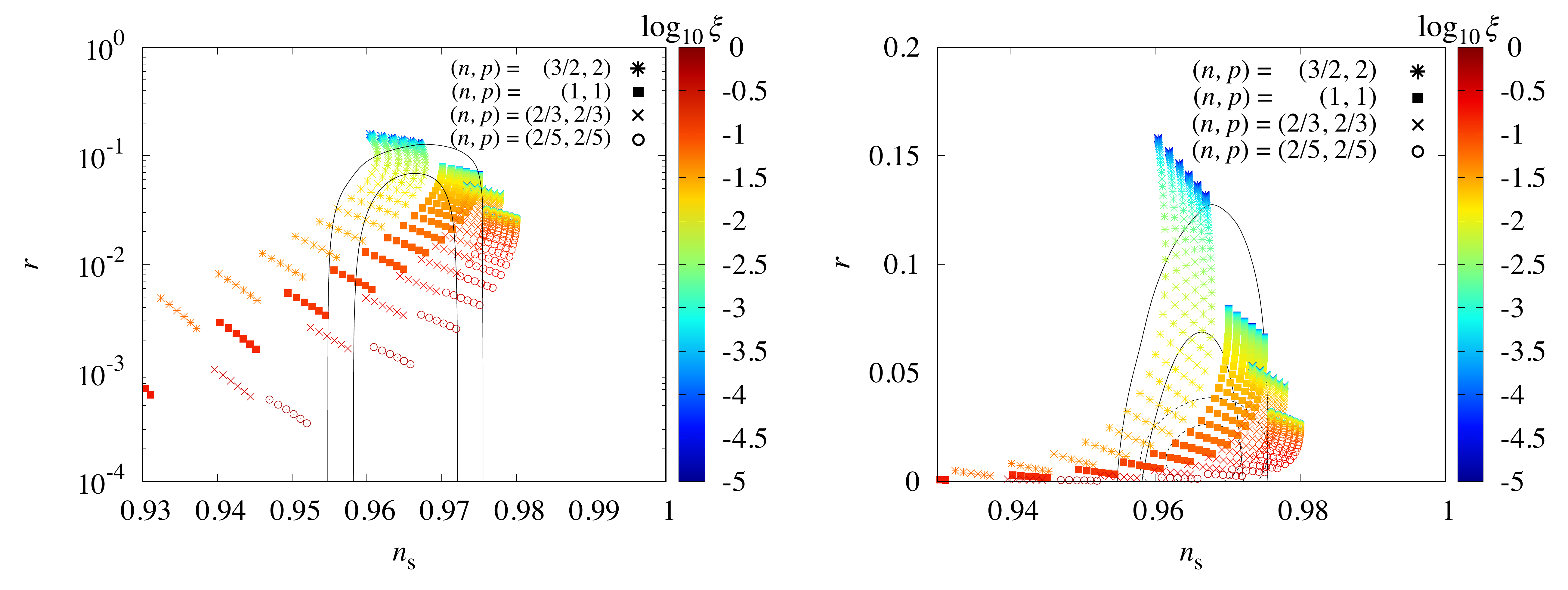}
\end{center}
\caption{\label{fig:F-dominant_chaotic}  Predictions of $n_s$ and $r$ for chaotic inflation with power-law ${\cal F}$ in the metric formulation with several values for $p$ and $n$,  all of which give the  ${\cal F}$-dominant type. In the left (right) panel, $r$ is shown with logarithmic (linear) scale and constraints from Planck~\cite{Planck:2018jri}  and Planck+BAO+BICEP/Keck~2018~\cite{BICEP:2021xfz} are also depicted with solid and dashed lines, respectively. When $r$ is shown in a log-scale (left panel), we only show the Planck constraints.}
\end{figure}

In Fig.~\ref{fig:F-dominant_chaotic}, the predictions for $n_s$ and $r$ for chaotic inflation with power-law ${\cal F}$ in the metric case are shown for several values of $n$ and $p$, all of which correspond to the $\cal F$-dominant type. We do not show the Palatini case since the results are quite similar to the metric one, particularly for the range of $\xi$ shown in the figure. One can see that the spectral index becomes smaller and the tensor-to-scalar ratio gets more suppressed as $\xi$ increases.  Although the minimally coupled chaotic inflation with any power-law index has already been excluded by Planck+BAO+BICEP/Keck~2018 data \cite{BICEP:2021xfz}, when the non-minimal coupling is introduced, the model can become viable,  when $p \lesssim 2$ by appropriately choosing the function ${\cal F}$.  As argued above, by increasing the value of $\xi$, the spectral index becomes more red-tilted, and the tensor-to-scalar ratio gets more suppressed in the $\cal F$-dominant type. Due to this feature, even when the minimally coupled version of this model is excluded by the fact that it predicts a larger $n_s$ and/or larger $r$, such models can be relaxed by introducing a non-minimal coupling of the $\cal F$-dominant type with some limited range of $\xi$. It should be also noted that, when $\xi$ is taken to be relatively large, $n_s$ gets too red-tilted, and then such cases are not compatible with observational constraints.

\subsubsection*{Example: Natural inflation with cosine-type $\cal{F}$}
In this model, the slow-roll parameter $\ep_J$ is given in \eqref{eq:eps_J_natural} and from Eq.~\eqref{eq:a_alpha_Fdom}, $a(\phi_{\rm ex})$ is related to $n$ as 
\begin{align}
a(\phi_{\rm ex})  = 2n -1\,.
\end{align}
Since $\phi_{\rm ex}$ is given by
\begin{align}
\label{eq:phi_ex_natural}
2\sin^2\left(\frac{\phi_{\rm ex}}{2f}\right) = \left[
\frac{1}{(2n -1)\xi}
\right]^{1/n}\,,
\end{align}
$\eta(\phi_{\rm ex})$ can be expressed as
\begin{align}
\label{eq:xi_eta_natural}
\eta(\phi_{\rm ex}) = -\left(\frac{\Mpl}{f}\right)^2n\left(
2\left\{(2n-1)\xi\right\}^{1/n} - 1
\right)\left[
1 + \frac{3}{4}\frac{n}{2n - 1} \left(\frac{\Mpl}{f}\right)^2\left(
2\left\{(2n-1)\xi\right\}^{1/n} - 1
\right)
\right]^{-1} \,.
\end{align}
For example,  when we take $n=2$ and $f = 7 \Mpl$, $\eta (\phi_{\rm ex})$ can be expressed, for the metric case, as 
\begin{equation}
\label{eq:eta_Fdom_natural}
\eta(\phi_{\rm ex}) = -2 \frac{2\sqrt{3\xi} - 1}{49\left(1 + \displaystyle\frac{1}{98}(2\sqrt{3\xi} - 1 )\right)}\,.
\end{equation}

\begin{figure}[ht]
\begin{center}
\includegraphics[width=16.5cm]{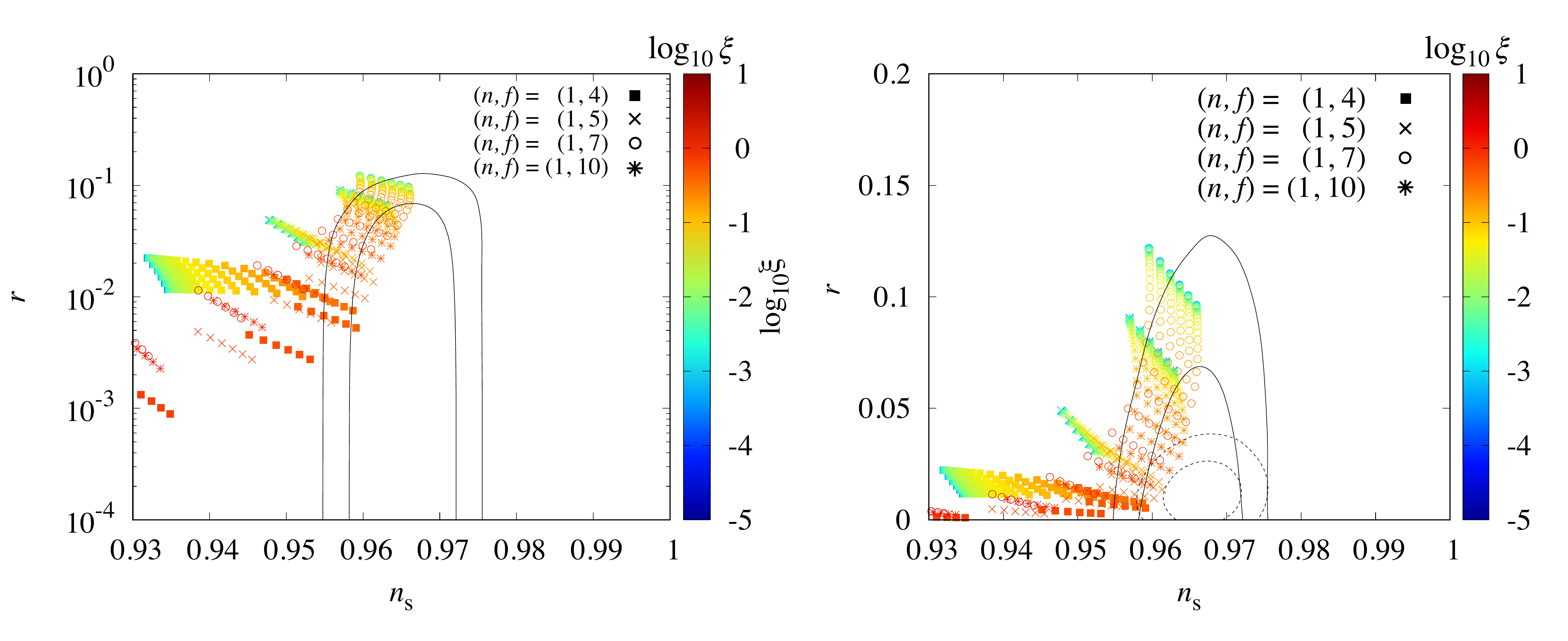}
\end{center}
\caption{\label{fig:F-dominant_natural}  Same as Fig.~\ref{fig:F-dominant_chaotic} but for natural inflation with cosine-type ${\cal F}$ for the metric case with $n=1$ corresponding to the ${\cal F}$-dominant type with several values of $f$. }
\end{figure}

In Fig.~\ref{fig:Fdom_eta_xi}, $\eta(\phi_{\rm ex})$ for natural inflation with cosine-type ${\cal F}$ is plotted as a function of $\xi$.  As in other cases such as the chaotic inflation with power-law $\cal F$, the numerical  and  analytic approximation given in Eq.~\eqref{eq:eta_Fdom_natural}  well agree for relatively  large values of $\xi$.

In Fig.~\ref{fig:F-dominant_natural}, the predictions of $n_s$ and $r$ are shown for the case of  $n=1$ with several values for $f$, all of which give the $\cal F$-dominant type. As in the case for chaotic inflation with power-law $\cal F$, the spectral index becomes more red-tilted and the tensor-to-scalar ratio gets more suppressed  for relatively large $\xi$ as seen from the figure. However, the behavior at small $\xi$ depends on the model parameters even among the $\cal F$-dominant type. For example, in the cases with $(n, f) = (1, 10 \Mpl)$ and $(n, f) = (1, 7 \Mpl)$ in Fig.~\ref{fig:F-dominant_natural}, the spectral index move from smaller to larger values as $\xi$ increases when $\xi$ is very small, however,  when $\xi$ becomes relatively large,  $n_s$ decreases as $\xi$ increases. Although the behavior at small $\xi$ generally depends on model parameters, the trend of $n_s$ and $r$ at relatively large $\xi$ can be understood from Eqs.~\eqref{eq:Fdom_r}, \eqref{eq:Fdom_ns} and \eqref{eq:eta_Fdom_ex}.  Since the minimally coupled version of natural inflation predicts $n_s$ comparable or smaller than the observational bound, and thus even if we introduce a non-minimal coupling, the $\cal F$-dominant type models would not be helpful much to alleviate the model except some limited range of $\xi$.

\subsubsection*{Example: Loop inflation with log-type $\cal{F}$}
The slow-roll parameter $\epsilon_J$ in this model is given by \eqref{eq:eps_J_loop} and $a(\phi_{\rm ex})$ can be written by $n$ as 
\begin{align}
a(\phi_{\rm ex}) = 2n  -1 \,,
\end{align}
which is the same as the case for natural inflation with cosine-type $\cal{F}$.  $\phi_{\rm ex}$ can be obtained from the extremum condition~\eqref{eq:extremum} as 
\begin{align}
\frac{\phi_{\rm ex}}{\Mpl} = \exp\left[
\frac{[(2n-1)\xi]^{-1/n} - 1}{a_h}
\right]\,.
\end{align}
By using this expression, $\eta (\phi_{\rm ex})$ can be given by 
\begin{align}
\label{eq:xi_eta_loop}
\eta(\phi_{\rm ex}) &= - a_h^2 n \left[ (2n-1) \xi \right]^{2/n}
\exp\left(
-2\frac{\left\{ (2n-1)\xi \right\}^{-1/n}-1}{a_h}
\right) \notag \\
& \qquad\qquad\qquad
  \times \left[
1 + \frac{3}{4}\frac{n}{2n - 1}a_h^2 \left[ (2n-1) \xi \right]^{2/n} 
\exp\left(
-2\frac{\left\{(2n-1)\xi\right\}^{-1/n}-1}{a_h}
\right)
\right]^{-1}\,.
\end{align}

In Fig.~\ref{fig:Fdom_eta_xi}, the value of $\eta$ for the case of $(n, a_h) = (2, 1)$ as a function of $\xi$ is shown along with that for other models. As in other models, $\eta$ becomes negatively large as $\xi$ increases. For relatively large $\xi$, $\eta$ is given, for the case of $(n, a_h) = (2, 1)$,  by 
\begin{equation}
\eta(\phi_{\rm ex}) = -\frac{ 6 \exp\big(-2(1/\sqrt{3\xi} - 1)\big)\xi}{1 + \displaystyle\frac{3}{2}\exp\big(-2(1/\sqrt{3\xi}-1)\big)\xi} \,.
\end{equation}
which matches well with numerically calculated one as seen from the figure. 

\begin{figure}[t]
\begin{center}
\includegraphics[width=16.5cm]{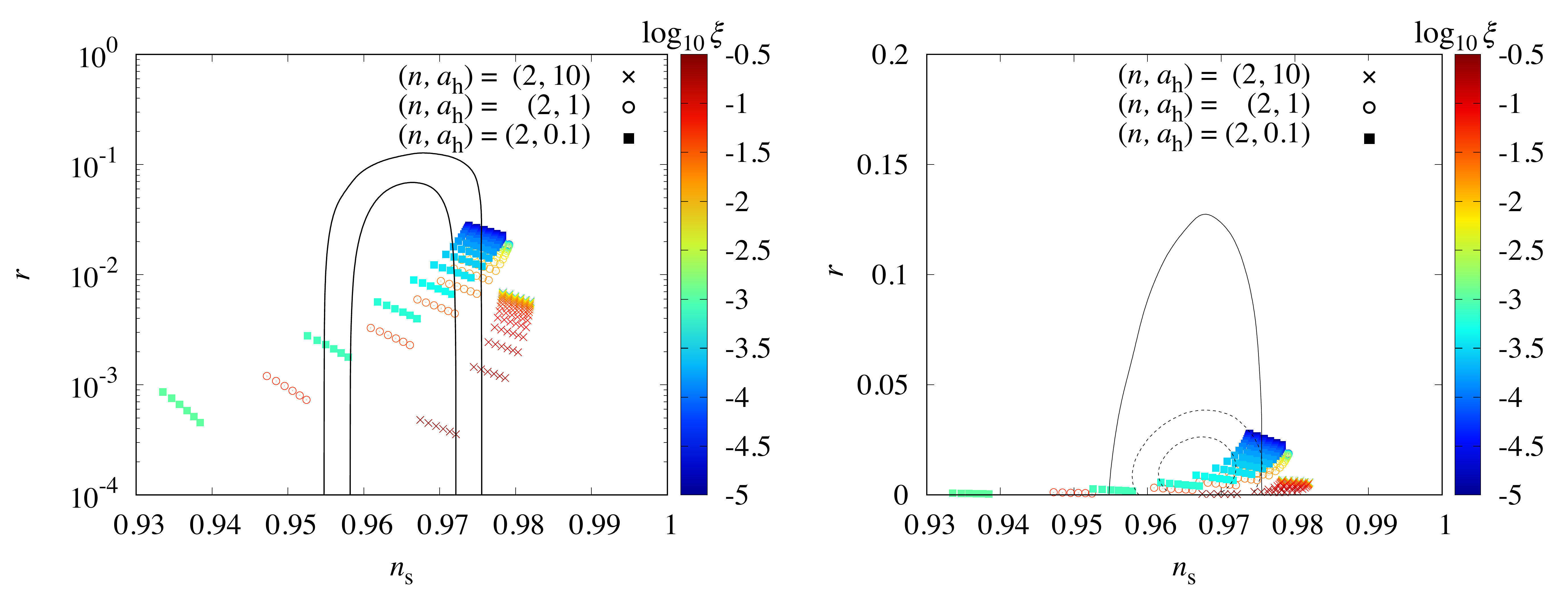}
\end{center}
\caption{\label{fig:F-dominant_loop}  Same as Fig.~\ref{fig:F-dominant_chaotic} but for loop inflation with log-type ${\cal F}$ for the metric case with $n=2$ corresponding to the ${\cal F}$-dominant type with several values of $a_h$.}
\end{figure}

In Fig,~\ref{fig:F-dominant_loop}, the predictions for  $n_s$ and $r$ are depicted for the metric case of $n=2$ with several values of $a_h$.  Since the minimally coupled counterpart of this model predicts a larger $n_s$ than the observational bound, by assuming the functional form for $\cal F$ corresponding to the $\cal F$-dominant type, $n_s$ can become smaller as $\xi$ increases. As mentioned in Sec.~\ref{sec:example_models}, the minimally coupled version of loop inflation with $a_h \lesssim 0.5$ are excluded by current data, however, such models become viable since the non-minimal coupling makes $n_s$ redder than that for the minimally coupled counterpart.  Like in chaotic inflation case, when the minimally coupled model  is excluded by the current data because $n_s$ is larger than the observational bound, the $\cal F$-dominant type model can relax such models. 

\section{Conclusion and Discussion \label{sec:conclusion}}

The power spectrum of primordial fluctuations are now well measured by cosmological observations such as CMB, large scale structure and so on. In particular, the spectral index $n_s$ and the tensor-to-scalar ratio $r$ are commonly used to test inflation models,  and the current constraints,  especially  from recent BICEP/Keck 2018 results in combination with Planck and BAO data, have excluded a lot of inflation models including chaotic inflation with any power-law index and natural inflation. However, when one introduces a non-minimal coupling to gravity, the predictions for $n_s$ and $r$ are modified,  and even if the minimally coupled version is excluded by observational constraint, such models can be relaxed.

In this paper, we have investigated this issue in a general setting. For this purpose, we have classified non-minimally coupled inflation models into three categories: the attractor, $V_J$-dominant and $\cal F$-dominant types by the functional forms of the Jordan frame potential $V_J$ and non-minimal coupling function $\cal F$. As we argued,  the predictions for $n_s$ and $r$ in each type exhibit different trends, and models in each category share those general features. 

In the attractor type,  the values of $n_s$ and $r$ approach to the fixed (attractor) values for the metric case  in the large  $\xi$ limit, given by the expressions~\eqref{eq:attractor_ns} and \eqref{eq:attractor_r},  which are well within the current observational constraint, and thus any models which satisfy the attractor type relation~\eqref{eq:attractor_type}  become viable.  Actually this kind of feature has  already been discussed for some particular models, and those can be categorized as the attractor type in our classification. In the Palatini case, $n_s$ approaches to the value given in Eq.~\eqref{eq:ns_attractor_palatini} which depends on the Jordan frame potential, whereas $r$ always gets suppressed as shown in Eq.~\eqref{eq:attractor_r_palatini}.  Therefore in this case, models particularly having been excluded by too large $r$ can be relaxed although the resultant spectral index depends on the Jordan frame potential, which may or may not fall onto the observational bounds.

The $V_J$-dominant type can be further divided into Case~(i) and (ii). For Case~(i), the predictions for $n_s$ and $r$ approach  the consistency relation given in Eq.~\eqref{eq:VJ_case1_consistency} independently from the forms of $V_J$ and ${\cal F}$. Unfortunately this consistency relation is already outside the observational constraints in the $n_s$--$r$ plane. In fact,  as illustrated for chaotic inflation with power-law $\cal F$ model in Fig.~\ref{fig:V_dominant_chaotic}, the trajectory from small to large $\xi$ to reach the consistency line is somewhat non-trivial, the model can become viable at some intermediate, but only for limited values of $\xi$. In general, this class of models would not help to relax (minimally coupled) inflation models. For Case~(ii), the general results are illustrated in Fig.~\ref{fig:V_dominant_case2}, from which one can see that when the Jordan frame slow-roll parameter $\eta_J$ is positive, such models are ruled out for $\alpha > 2/3$. Even when $\eta_J$ is negative, although some parameter space may be relaxed such as in natural inflation case, this type generally would not be helpful to alleviate inflation models, which we demonstrated for some explicit ones.

In the $\cal F$-dominant type, the general prediction would be that $n_s$ gets more red-tilted and $r$ is more suppressed as $\xi$ increases. Therefore, for minimally coupled models which predict a larger $n_s$ and/or a larger $r$ than the observational constraints, the introduction of a non-minimal coupling would help to relax the model by appropriately choosing the value of $\xi$. One of such examples is the chaotic inflation, shown in Fig.~\ref{fig:F-dominant_chaotic}, in which several models become viable by assuming some range of $\xi$. Loop inflation with log-type ${\cal F}$ can also be saved since the minimally coupled counterpart gives a bit bluer $n_s$ compared to the observational bound. However natural inflation would not be relaxed in broad parameter space since the minimally coupled version of this model tends to predict too red-tilted spectrum. The ${\cal F}$-dominant type can be helpful to relax inflation models depending on the predictions for $n_s$ and $r$ in the original (minimally coupled) model. 

Since cosmological observations are now very precise and would be more accurate in the future, constraints on inflation models will become severer and more models would be excluded. However, by extending the model framework by introducing a non-minimal coupling to gravity, which is studied in this paper, the predictions for $n_s$ and $r$ are modified and some models can be viable in the extended framework. Our study would  give a more opportunity to model building or designing an inflation model consistent with observational bounds in the framework of non-minimal coupling to gravity.

\section*{Acknowledgements}
The authors thank Tommi~Tenkanen for the collaboration at early stage of this work.
The work of T.T. was supported by JSPS KAKENHI Grant Number 17H01131,  19K03874 and MEXT KAKENHI Grant Number 19H05110.

\clearpage 

\bibliography{non-minimal_inflation}

\end{document}